\documentclass[10pt]{revtex4}
\usepackage[pdftex]{graphicx}
\usepackage{epstopdf}
\usepackage{amssymb}
\usepackage{latexsym}
\usepackage{epsfig}
\usepackage{float}
\usepackage{graphicx,epsfig,color }
\usepackage{graphicx}
\usepackage{subfigure}
\usepackage{amsthm}
\usepackage{amsmath}

\usepackage{hyperref}

\begin{document}
\title{Yang-Mills black holes in Quasitopological gravity}
\author{Fatemeh Naeimipour\footnote{sara.naeimipour1367@gmail.com}, Behrouz Mirza\footnote{b.mirza@iut.ac.ir}, Fatemeh Masoumi Jahromi\footnote{fatemehmasoumi@ph.iut.ac.ir}}
\address{Department of Physics, Isfahan University of Technology, Isfahan 84156-83111, Iran}

\begin{abstract}
In this paper, we formulate two new classes of black hole solutions in higher curvature quartic quasitopological gravity with nonabelian Yang-Mills theory. At first step, we consider the $SO(n)$ and $SO(n-1,1)$ semisimple gauge groups. We obtain the analytic quartic quasitopological Yang-Mills black hole solutions. Real solutions are only accessible for the positive value of the redefined quartic quasitopological gravity coefficient, $\mu_{4}$. These solutions have a finite value and an essential singularity at the origin, $r=0$ for space dimension higher than $8$. We also probe the thermodynamic and critical behavior of the quasitopological Yang-Mills black hole. The obtained solutions may be thermally stable only in the canonical ensemble. They may also show a first order phase transition from a small to a large black hole. In the second step, we obtain the pure quasitopological Yang-Mills black hole solutions. For the positive cosmological constant and the space dimensions greater than eight, the pure quasitopological Yang-Mills solutions have the ability to produce both the asymptotically AdS and dS black holes for respectively the negative and positive constant curvatures, $k=-1$ and $k=+1$. This is unlike the quasitopological Yang-Mills theory which can lead to just the asymptotically dS solutions for $\Lambda>0$. The pure quasitopological Yang-Mills black hole is not thermally stable.   
\end{abstract}


\maketitle

\section{Introduction}
Einstein's gravity as a relatively weak theory may be the only appropriate model in four dimension. For higher dimensions, Einstein's equations are not the most perfect ones which can satisfy the Einstein’s assumptions. Higher-dimensional spacetime as a requirement of the string theory also plays an important role in the AdS/CFT correspondence which makes a relation between the $n$-dimensional conformal field theories and the $(n+1)$-dimensional anti-de Sitter (AdS) black holes \cite{Aha}. Therefore, we can extend our study to the generalized gravities with the higher-order derivative terms that lead to the second order field equations.\\ Quasitopological gravity is one of the candidates of the modified theories having the ability to resolve the Einstein's equation defects \cite{Myers1,Oliva}. From the viewpoint of the AdS/CFT correspondence, quasitopological gravity can cause a broader class of four (and higher)-dimensional CFT's which includes three (or higher) independent parameters relating the central charges of the conformal field theories with the coupling parameters of the gravitational spacetimes\cite{Myers1,Lemos,Lemos1,Mann1,Deh1}. On the other hand, by choosing some special
constraints on the coupling constants of the quasitopological gravity, the causality bound on the CFT's can be respected \cite{Myers2}. This gravity has also priority over the Lovelock theory \cite{LoveLock}. As the Euler density terms of the quasitopological gravity are not true topological, this gravity contributes to the equations of motion in higher dimensions. For example, the fourth-order Lovelock theory contributes to the equations of motions only in nine and higher dimensions. This is while that the quartic quasitopological gravity contributes to the field equations for space dimensions $n\geq 4$, except for $n=7$ \cite{Myers1}. Therefore, this leads to a wide class of dual CFT's for lower dimensions \cite{Myers1}. Until now, some investigations of the quasitopological black hole solutions have been done \cite{Oliva,Myers1,Deha,Dehb,Brenna1}. Two classes of the uncharged and charged quartic quasitopological black holes with the linear Maxwell theory have been studied in respectively Refs. \cite{Bazr1} and \cite{Naeimi3}. Rotating black branes and magnetic branes in the presence of the quasitopological gravity have been respectively probed in Refs. \cite{Naeimi1} and \cite{Naeimi2}. \\
The nonabelian Yang-Mills theory is a generalization of the abelin Maxwell theory which is more common in the gauged supergravity AdS theories. The idea of considering Yang-Mills theory with gravity was first studied in Ref. \cite{Bartnik}. By using a numeric method in four dimension, the authors could achieve a class of asymptotically flat spherically symmetric solitonic
solution in the presence of a SU(2) Einstein-Yang-Mills field. The first black hole solution in the presence of the Einstein-Yang-Mills theory was constructed in Ref. \cite{Yasskin}. Einstein-Yang-Mills solutions in the presence of the cosmological constant have been checked out in Refs.\cite{Torii,Bjoraker}. Some other studies of the Yang-Mills theory are in Refs. \cite{Volkov,Brihaye,Wu,Mazharimousavi,Bostani}. In this paper, we are eager to use the Wu-Yang ansatz \cite{Wu} and obtain an analytic quasitopological Yang-Mills black hole solution in the presence of the $SO(n)$ and $SO(n-1,1)$ gauge groups. \\
Recently, the idea of pure Lovelock gravity has attracted a lot of attention. It originates from the idea of constructing a black string/brane in Lovelock gravity. Unlike the pure Einstein gravity which can lead to a black string/brane by adding flat directions to a vacuum black hole solution, it is not possible to obtain a black string/brane in the Lovelock theory. Pure Lovelock gravity with just one Euler density term was introduced with the aim of solving this problem \cite{Kastor1}. Black holes in pure Lovelock gravity have been investigated in Ref. \cite{Cai0}. In this study, an asymptotically AdS black hole is accessible only for the constant curvature $k=-1$ which is related to the hyperbolic angular coordinates. This is while that in general relativity, an AdS black hole can be described by all, $k=-1,0,+1$, where the constant curvatures $k=0,+1$ describe the flat and spherical angular coordinates. In Ref. \cite{Naresh1}, the authors have proved that the pure Lovelock black hole in the dimension, $d=3N+1$, is stable. Some studies about the quasi-local energy and ADM mass of the pure Lovelock gravity are in Ref. \cite{Jani}. Thermodynamic extended phase space and $P-V$ criticality of the black holes in pure Lovelock gravity
have been also probed \cite{Estrada}. Based on the advantages of the quasitopological gravity over the Lovelock theory as mentioned, now, we are willing to obtain the solutions of the pure quasitopological Yang-Mills black hole in the second part of this paper.\\
The outline of this paper is as follows: In Sec.\ref{Field}, we obtain the black hole solutions of the $(n+1)$-dimensional quartic quasitopological gravity coupled to the Yang-Mills theory and then discuss the physical properties of the solutions. Then, we probe the thermodynamic behaviors of the Yang-Mills quartic quasitopological black hole and study the thermal stability in Sec.\ref{thermoY}. We also study the critical behavior of the Yang-Mills quasitopological black hole in Sec.\ref{critical1}. In the second part of this paper, we obtain the pure quasitopological Yang-Mills black hole solution and investigate the physical properties in Sec.\ref{pure}. We also study the thermodynamic behaviors and thermal stability of this black hole in Sec. \ref{thermoP} and finally, we have a conclusion of the whole paper in Sec.\ref{con}.
 %%%%%%%%%%%%%%%%%%%%%%%%%%%%%%%%%%%%%%%%%%%%%%%%%%%%
\section{General structure of the Quasitopological Gravity Coupled to the Yang-Mills theory}\label{Field}
Higher dimensional quasitopological gravity with Maxwell theory has been defined in Ref. \cite{Naeimi3}. In this section, we first introduce the quasitopological gravity and the nonabelian Yang-Mills theory and then obtain the quasitopological Yang-Mills black hole solutions. We consider a N-parameters gauge group $\mathcal{G}$ with the structure constants $C^{a}_{bc}$ which have the definitions
\begin{eqnarray}
\gamma_{ab}\equiv-\frac{\Gamma_{ab}}{|\mathrm {det} \Gamma_{ab}|^{1/N}}\,\,\,\rm{and}\,\,\,  \Gamma_{ab}\equiv C_{ad}^{c}C_{bc}^{d},
\end{eqnarray}
and the indices $a$, $b$, $c$ goes from 1 to N. The $(n+1)$-dimensional action for the quasitopological gravity with the nonabelian Yang-Mills theory is followed by 
\begin{equation}\label{Act1}
I_{\rm{bulk}}=\frac{1}{16\pi}\int{d^{n+1}x\sqrt{-g}\big\{-2\Lambda+{\mathcal L}_1+\hat{\mu}_{2} {\mathcal L}_2+\hat{\mu}_{3} {\mathcal L}_3+\hat{\mu}_{4}{\mathcal L}_4-\gamma_{ab}F_{\mu\nu}^{(a)}F^{(b)\mu\nu}\big\}},
\end{equation}
where $\Lambda$ is the cosmological constant and $\hat{\mu}_2$, $\hat{\mu}_3$ and $\hat{\mu}_4$ are the coefficients of the quasitopological gravity. ${\mathcal L}_1=R$, ${\mathcal L}_2$, ${\mathcal L}_3$ and ${{\mathcal L}_4}$ are respectively, Einstein-Hilbert, the second-order Lovelock (Gauss-Bonnet term), the cubic and quartic quasitopological gravities which are defined as  
\begin{eqnarray}
{\mathcal L}_2&=& R_{abcd}R^{abcd}-4R_{ab}R^{ab}+R^2,
\end{eqnarray}
\begin{eqnarray}\label{quasi3}
{\mathcal{L}_3}&=&R_a{{}^c{{}_b{{}^d}}}R_c{{}^e{{}_d{{}^f}}}R_e{{}^a{{}_f{{}^b}}}+\frac{1}{8(2n-1)(n-3)} \big(b_{1}R_{abcd}R^{abcd}R+b_{2}R_{abcd}R^{abc}{{}_e}R^{de}\nonumber\\
&&+b_{3}R_{abcd}R^{ac}R^{bd}+b_{4}R_a{{}^b}R_b{{}^c}R_{c}{{}^a}+b_{5}R_a{{}^b}R_b{{}^a}R +b_{6}R^3\big),
\end{eqnarray}
\begin{eqnarray}\label{quasi4}
{\mathcal{L}_4}&=& c_{1}R_{abcd}R^{cdef}R^{hg}{{}_{ef}}R_{hg}{{}^{ab}}+c_{2}R_{abcd}R^{abcd}R_{ef}{{}^{ef}}+c_{3}RR_{ab}R^{ac}R_c{{}^b}+c_{4}(R_{abcd}R^{abcd})^2\nonumber\\
&&+c_{5}R_{ab}R^{ac}R_{cd}R^{db}+c_{6}RR_{abcd}R^{ac}R^{db}+c_{7}R_{abcd}R^{ac}R^{be}R^d{{}_e}+c_{8}R_{abcd}R^{acef}R^b{{}_e}R^d{{}_f}\nonumber\\
&&+c_{9}R_{abcd}R^{ac}R_{ef}R^{bedf}+c_{10}R^4+c_{11}R^2 R_{abcd}R^{abcd}+c_{12}R^2 R_{ab}R^{ab}\nonumber\\
&&+c_{13}R_{abcd}R^{abef}R_{ef}{{}^c{{}_g}}R^{dg}+c_{14}R_{abcd}R^{aecf}R_{gehf}R^{gbhd},
\end{eqnarray}
where we have written the coefficients $c_{i}$'s in the appendix \eqref{app1}.
The Yang-Mills gauge field tensor is of the form
\begin{eqnarray}
F_{\mu\nu}^{(a)}=\partial_{\mu}A_{\nu}^{(a)}-\partial_{\nu}A_{\mu}^{(a)}+\frac{1}{e}C^{a}_{bc}A_{\mu}^{(b)}A_{\nu}^{(c)}, 
\end{eqnarray}
where $e$ is a coupling constant and $A_{\mu}^{(a)}$'s are the gauge potentials. We use the metric
\begin{eqnarray}\label{metric}
ds^2=-f(r)dt^2+\frac{dr^2}{f(r)}+r^2[d\theta^2+k^{-1}\mathrm{sin}^{2} (\sqrt{k}\theta)d\Omega_{k,n-2}^2],
\end{eqnarray}
where $d\Omega_{k,n-2}^2$ represents the metric of a unit $(n-2)$-sphere with constant curvatures $k=-1,1$ that are respectively related to the hyperbolic and spherical angular coordinates. We introduce the following coordinates to write the Yang-Mills potentials
\begin{eqnarray}
x_{1}&=&\frac{r}{\sqrt{k}}\, \mathrm{sin}(\sqrt{k}\,\theta)\,\Pi_{j=1}^{n-2}\,\mathrm{sin}\,\phi_{j},\nonumber\\
x_{l}&=&\frac{r}{\sqrt{k}}\, \mathrm{sin}(\sqrt{k}\,\theta)\,\mathrm{cos}\,\phi_{n-l}\,\Pi_{j=1}^{n-l-1}\,\mathrm{sin}(\phi_{j})\,\,\,\,,\,\,\,l=2, ..., n-1\nonumber\\
x_{n}&=& r\, \mathrm{cos}\,(\sqrt{k}\,\theta),
\end{eqnarray}
where $k=-1,+1$ are the constant curvatures. By the Wu-Yang ansatz \cite{Wu}, the gauge potentials can be derived
\begin{eqnarray}\label{potential}
A^{(a)}&=&\frac{e}{r^2}(x_{l}dx_{n}-x_{n}dx_{l})\,\,\,\mathrm{for}\,\,\, a=l=1,...,n-1,\nonumber\\
A^{(b)}&=&\frac{e}{r^2}(x_{l}dx_{j}-x_{j}dx_{l})\,\,\, \mathrm{for}\,\,\,\,b=n,...,n(n-1)/2,\,\,\,l=1,...,n-2,\,\,\,j=2,...,n-1,\mathrm{and}\,l<j. 
\end{eqnarray}
For a better understanding, we have written the gauge potentials of the gauge groups $SO(3)$, $SO(2,1)$, $SO(4)$ and $SO(3,1)$ in the appendix\eqref{app2}. 
The gauge potentials have the Lie algebra of $SO(n-1,1)$ and $SO(n)$.
To obtain the gravitational field equation, we should vary the action \eqref{Act1} with respect to the metric $g_{\mu\nu}$. If we use the gauge potentials \eqref{potential} and redefine the coefficients in the following way 
\begin{eqnarray}
\mu_{2}&\equiv&(n-2)(n-3)\hat{\mu}_{2},\nonumber\\
\mu_{3}&\equiv&\frac{(n-2)(n-5)(3n^2-9n+4)}{8(2n-1)}\hat{\mu}_{3},\nonumber\\
\mu_{4}&\equiv& n(n-1)(n-3)(n-7)(n-2)^2(n^5-15n^4+72n^3-156n^2+150n-42)\hat{\mu}_{4},
\end{eqnarray} 
then the fourth-order gravitational field equation is obtained as 
\begin{eqnarray}\label{EEE1}
\mu_{4} \Psi^4+\mu_{3}\Psi^3+ \mu_{2}\Psi^2+\Psi+\zeta=0,
\end{eqnarray}
where $\Psi(r)=[k-f(r)]/r^2$ and 
\begin{eqnarray}\label{zeta1}
\zeta&=&\left\{
\begin{array}{ll}
$$-\frac{2\Lambda}{n(n-1)}-\frac{m}{r^n}-\frac{(n-2)e^2}{(n-4)r^{4}}$$,\quad \quad\quad\quad \  \ {n>4,}\quad &  \\ \\
$$-\frac{\Lambda}{6}-\frac{m}{r^4}-\frac{2 e^2}{r^4}\mathrm{ln}(\frac{r}{r_{0}})$$. \quad\quad \quad\quad\quad\quad\,\,\,\,{n=4,}\quad &
\end{array}
\right.
\end{eqnarray}
that we choose $r_{0}=1$ for simplification. In the above relation, $m$ is an integration constant which is interpreted as the mass of the black hole. The quasitopological Yang-Mills black hole solutions of Eq.\eqref{EEE1} is obtained as follows
\begin{eqnarray}\label{func11}
f(r)=k-r^2\times\left\{
\begin{array}{ll}
-\frac{\mu_{3}}{4\mu_{4}}+$$\frac{-W+\sqrt{-(3A+2y-\frac{2B}{W})}}{2}$$,\quad \quad\quad\quad \  \ {\mu_{4}>0,}\quad &  \\ \\
-\frac{\mu_{3}}{4\mu_{4}}+$$\frac{W-\sqrt{-(3A+2y+\frac{2B}{W})}}{2}$$, \quad \quad\quad\quad\quad\,{\mu_{4}<0,}\quad &
\end{array}
\right.
\end{eqnarray} 
where, for simplicity we have used the following definitions
\begin{equation}\label{yy}
W=\sqrt{A+2y}\,\,\,\,\,,\,\,\,\,\,A=-\frac{3\mu_{3}^{2}}{8\mu_{4}^2}+\frac{\mu_{2}}{\mu_{4}}\,\,\,\,\,\,\,\,\,\,\,,\,\,\,\,\,\,y=\left\{
\begin{array}{ll}
$$-\frac{5}{6}A+U-\frac{P}{3U}$$,\quad \quad\quad\quad \  \ {U\neq 0,}\quad &  \\ \\
$$-\frac{5}{6}A+U-\sqrt[3]{H}$$, \quad \quad\quad\quad{U=0,}\quad &
\end{array}
\right.
\end{equation}
\begin{eqnarray}\label{UU}
U=\bigg(-\frac{H}{2}\pm\sqrt{\frac{H^2}{4}+\frac{P^3}{27}}\,\bigg)^{\frac{1}{3}},\,\,H=-\frac{A^3}{108}+\frac{A C}{3}-\frac{B^2}{8}\,\,\,\,\,,\,\,\,\,\,\,\,
P=-\frac{A^2}{12}-C,
\end{eqnarray}
\begin{eqnarray}\label{eq11}
B=\frac{\mu_{3}^3}{8\mu_{4}^3}-\frac{\mu_{2}\mu_{3}}{2\mu_{4}^2}+\frac{1}{\mu_{4}}\,\,\,\,\,\,\,,\,\,\,\,\,\, C=-\frac{3\mu_{3}^4}{256\mu_{4}^4}+\frac{\mu_{2}\mu_{3}^2}{16\mu_{4}^3}-\frac{\mu_{3}}{4\mu_{4}^2}+\frac{\zeta}{\mu_{4}}.
\end{eqnarray}
It is clear from Eq.\eqref{func11} that the solutions are divided into two categories for $\mu_{4}>0$ and $\mu_{4}<0$. For small $r$, the parameter $\zeta$ in Eq.\eqref{zeta1} takes a negative large value which can enlarge the fourth term of the parameter $C$ in Eq. \eqref{eq11} and so the parameter $P$ in Eq. \eqref{UU}. For $\mu_{4}<0$, this leads to a negative large value for the parameter $P$ in Eq.\eqref{UU} which produces an imaginary solution for the parameter $U$. Thus, we do not consider the solutions with $\mu_{4}<0$ because they are imaginary for small $r$.\\  
Now, we would like to investigate the physical properties of the obtained solutions in two cases, $\hat{\mu}_{i}=0$ and $\hat{\mu}_{i}\neq0$, which $i=2,3$ .
\subsection{The $\hat{\mu}_{2}=0$ and $\hat{\mu}_{3}=0$ case}\label{mu2=0}
\begin{figure}
\centering
\subfigure[$\hat{\mu}_{4}=0.005$]{\includegraphics[scale=0.27]{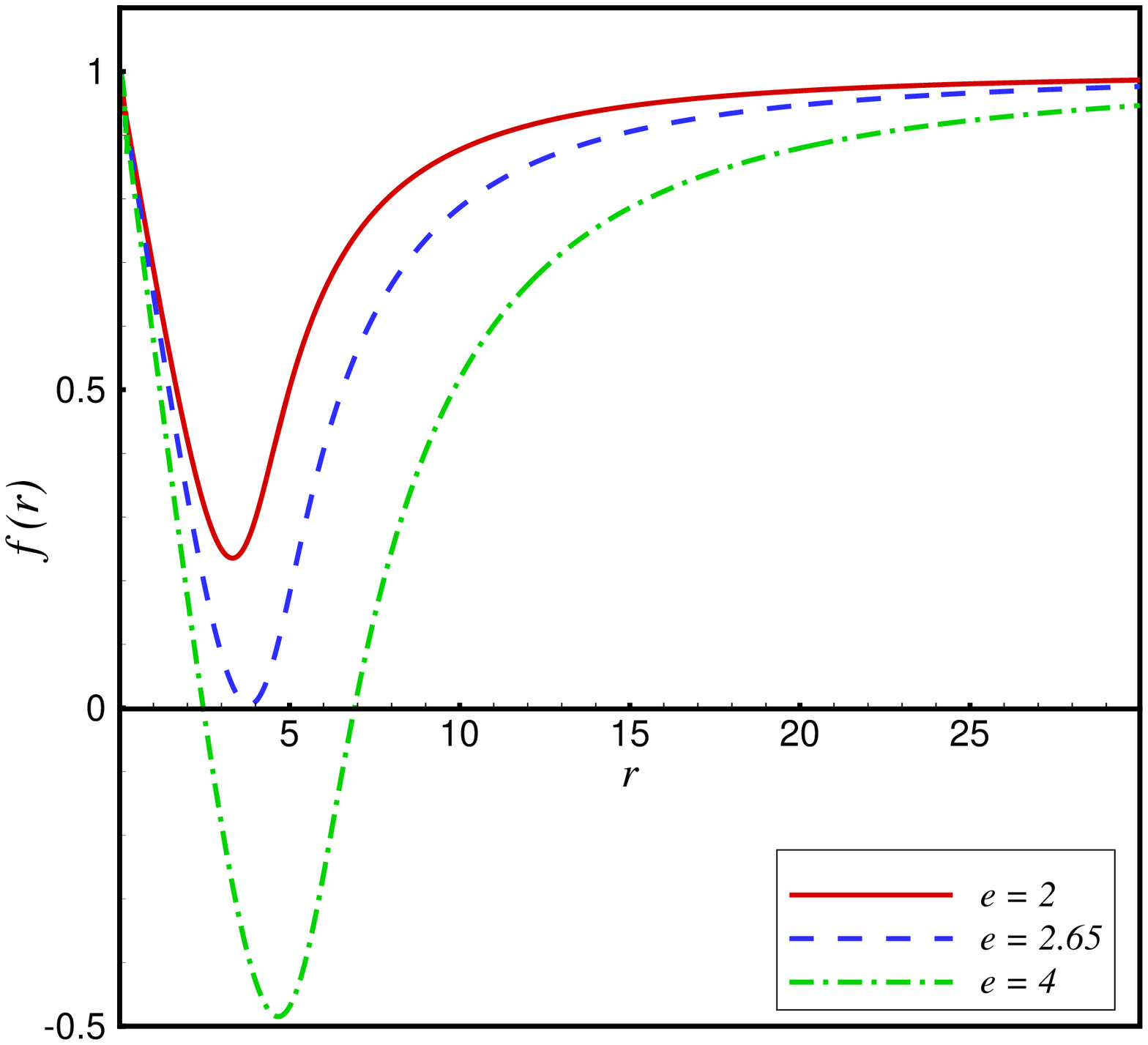}\label{Fig1a}}\hspace*{.2cm}
\subfigure[$\hat{\mu}_{4}=5\times 10^{-7}$]{\includegraphics[scale=0.27]{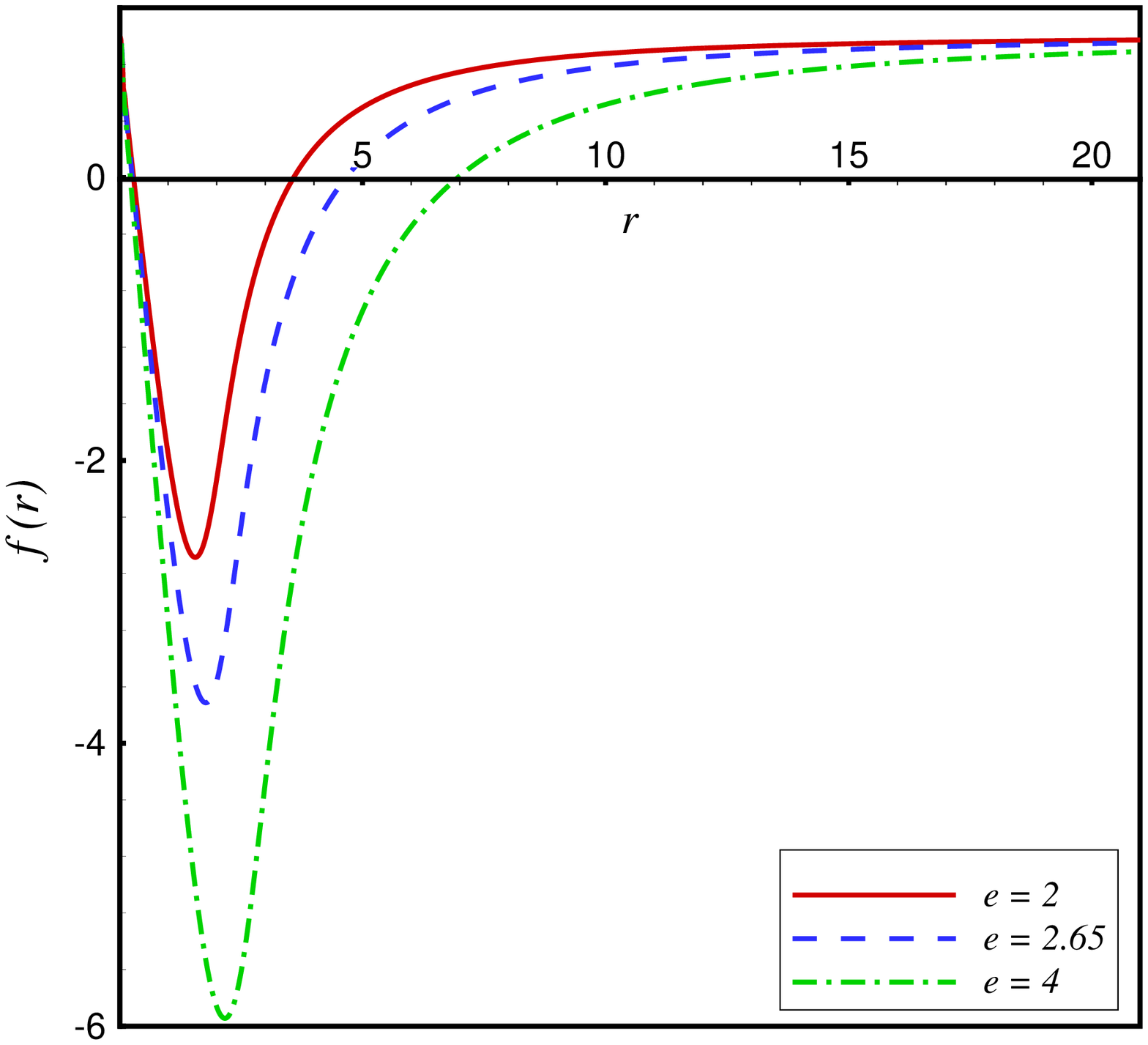}\label{Fig1b}}\hspace*{.2cm}
\subfigure[$f_{\mathrm{Einsten-Yang-Mills}}$]{\includegraphics[scale=0.27]{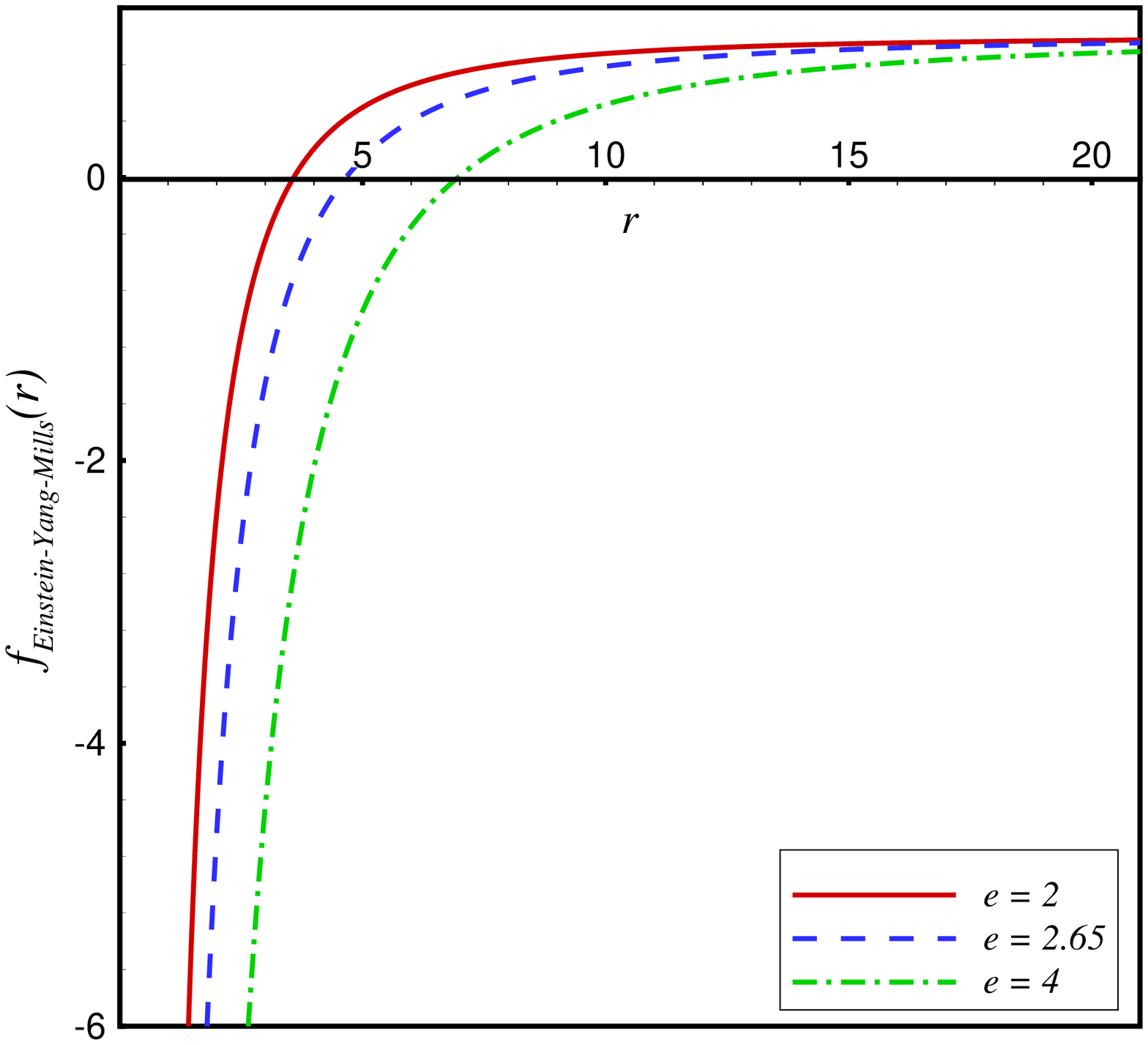}\label{Fig1c}}\caption{ $f(r)$ with respect to $r$ for different values of the Yang-Mills charge $e$ with $\mu_{2}=\mu_{3}=0$, $k=1$, $n=5$, $m=3$ and $\Lambda=0$.}\label{Fig1}
\end{figure}

In an attempt to get a simple expression for the obtained solution $f(r)$, we consider the special cases $\hat{\mu}_{2}=0$ and $\hat{\mu}_{3}=0$. In this case, the quartic quasitopological Yang-Mills solution gets the form
\begin{eqnarray}\label{fmu20}
f(r)&=& k-\frac{r^2}{2}\bigg[\mp \sqrt{2\big(\frac{1}{16\mu_{4}^2}+\sqrt{\Delta}\big)^{\frac{1}{3}}+\frac{2\xi}{3\mu_{4}\big(\frac{1}{16\mu_{4}^2}+\sqrt{\Delta}\big)^{\frac{1}{3}}}}
\pm\bigg(-2\bigg(\frac{1}{16\mu_{4}^2}+\sqrt{\Delta}\bigg)^{\frac{1}{3}}-\frac{2\xi}{3\mu_{4}\bigg(\frac{1}{16\mu_{4}^2}+\sqrt{\Delta}\bigg)^{\frac{1}{3}}}\nonumber\\
&&\pm\frac{2}{\mu_{4}}\bigg(2\bigg(\frac{1}{16\mu_{4}^2}+\sqrt{\Delta}\bigg)^{\frac{1}{3}}+\frac{2\xi}{3\mu_{4}(\frac{1}{16\mu_{4}^2}+\sqrt{\Delta})^{\frac{1}{3}}}\bigg)^{-\frac{1}{2}}\bigg)^{\frac{1}{2}}\bigg],
\end{eqnarray}
where 
\begin{eqnarray}\label{delta}
\Delta&=&\frac{1}{256\mu_{4}^4}-\frac{\zeta^3}{27\mu_{4}^3}.
\end{eqnarray}
The obtained signs $(-,+,+)$ and $(+,-,-)$ in Eq.\eqref{fmu20} are respectively corresponded to $\mu_{4}>0$ and $\mu_{4}<0$. The expansion of $f(r)$ around $\mu_{4}\rightarrow 0$ reduces to
\begin{eqnarray}
f(r)=k+\zeta r^2+\mu_{4}\zeta^4 r^2+4\mu_{4}^2\zeta^7 r^2+\mathcal{O}((\mu_{4})^{8/3}),
\end{eqnarray}
where $\zeta$ was defined in Eq. \eqref{zeta1}. This relation shows the Einstein-Yang-Mills solutions with some correction terms proportional to $\mu_{4}$. \\
We also check the solution $f(r)$ at the origin, $r=0$. As the term $m/r^n$ becomes so large at $r=0$ in Eqs. \eqref{fmu20} and \eqref{delta}, we should choose $\mu_{4}>0$ in order to have a real solution. Therefore, if we expand $f(r)$ near the origin for the positive finite values of $\mu_{4}$ and $n\neq 4$, it is obtained as follows 
\begin{eqnarray}
f(r)=k-\bigg(\frac{m}{\mu_{4}}\bigg)^{\frac{1}{4}}r^{\frac{8-n}{4}}-\frac{(n-2)e^2}{4(n-4)m}\times\bigg(\frac{m}{\mu_{4}}\bigg)^{\frac{1}{4}}r^{\frac{3n-8}{4}}+\mathcal{O}(r).
\end{eqnarray}
This shows that the quasitopological gravity has the ability to provide a finite value for the laps function at the origin, for $n\leq 8$. This is unlike the Einstein gravity which makes a divegent value at this point. If we calculate the Kretchmann scalar for the positive finite value of $\mu_{4}$ at the limit $r\rightarrow 0$, we get to 
\begin{eqnarray}
R_{abcd}R^{abcd} \propto\sqrt{\frac{m}{\mu_{4}}}r^{-\frac{n}{2}}.
\end{eqnarray}
This manifests a divergence at $r=0$ which demonstrates that there is an essential singularity located at the origin.\\
In order to have a better look of the obtained solution, we have plotted $f(r)$ versus $r$ for $\hat{\mu}_4=0.005$ in Fig.\ref{Fig1a}, for $\hat{\mu}_{4}=5\times 10^{-7}$ in Fig.\ref{Fig1b} and the Einstein-Yang-Mills solution in Fig.\ref{Fig1c}. We can observe that for the mentioned parameters in Fig.\ref{Fig1a}, there is an extreme black hole with the Yang-Mills charge, $e_{\mathrm{ext}}=2.65$. For $e>e_{\mathrm{ext}}$, there is a black hole with two horizons, while for $e<e_{\mathrm{ext}}$, there is a naked singularity. Fig.\ref{Fig1a} also shows that the function $f(r)$ in quasitopological gravity has a finite value at $r=0$. By decreasing the value of $\hat{\mu}_{4}$ in Fig.\ref{Fig1b}, $f(r)$ increases near the origin while in the case of the Einstein-Yang-Mills solution in Fig.\ref{Fig1c}, it diverges as $r \rightarrow 0$. So unlike the Einstein's gravity, the quasitopological gravity is successful to create a finite value for the laps function at $r=0$. The figures also show that for $r\rightarrow\infty$, the function $f(r)$ has a similar behavior in both Einstein-Yang-Mills and quasitopological-Yang-Mills theories. So, we can deduce that the effect of the quasitopological gravity will be removed at the infinity. In Figs.\ref{Fig1b} and \ref{Fig1c}, the radius of the event horizon $r_{+}$ (which is the root of the equation $f(r_{+})=0$) increases as the value of $e$ increases. 
\subsection{The $\hat{\mu}_{i}\neq0$ case}\label{mu2n0}
\begin{figure}
\center
\includegraphics[scale=0.5]{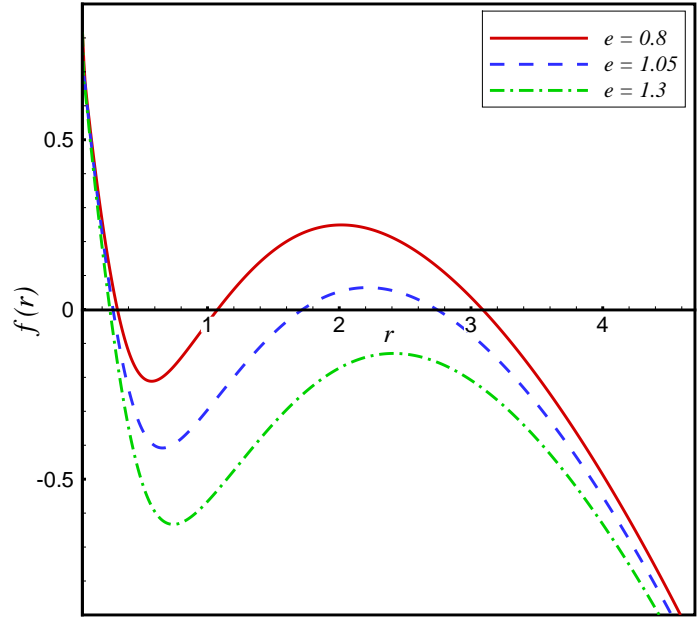}
\caption{\small{$f(r)$ with respect to $r$ for different values of $e$ with $\hat{\mu}_{2}=0.3$, $\hat{\mu}_{3}=0$, $\hat{\mu}_{4}=10^{-7}$, $k=1$, $n=5$, $m=1$ and $\Lambda=1$. } \label{Fig2}}
\end{figure}

\begin{figure}
\center
\includegraphics[scale=0.5]{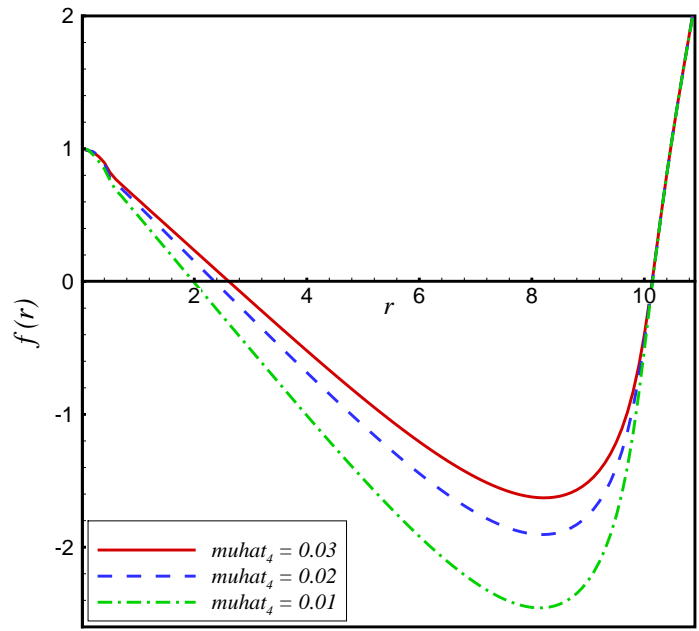}
\caption{\small{$f(r)$ with respect to $r$ for different values of $\hat{\mu_{4}}$ with $\hat{\mu}_{2}=-0.9$, $\hat{\mu}_{3}=-0.06$, $k=1$, $n=6$, $m=10$, $e=20$ and $\Lambda=-1$.} \label{Fig3}}
\end{figure}

\begin{figure}
\centering
\subfigure[$\hat{\mu}_{2}=-0.9$, $\hat{\mu}_{3}=-0.6$, $\hat{\mu}_{4}=0.005$]{\includegraphics[scale=0.27]{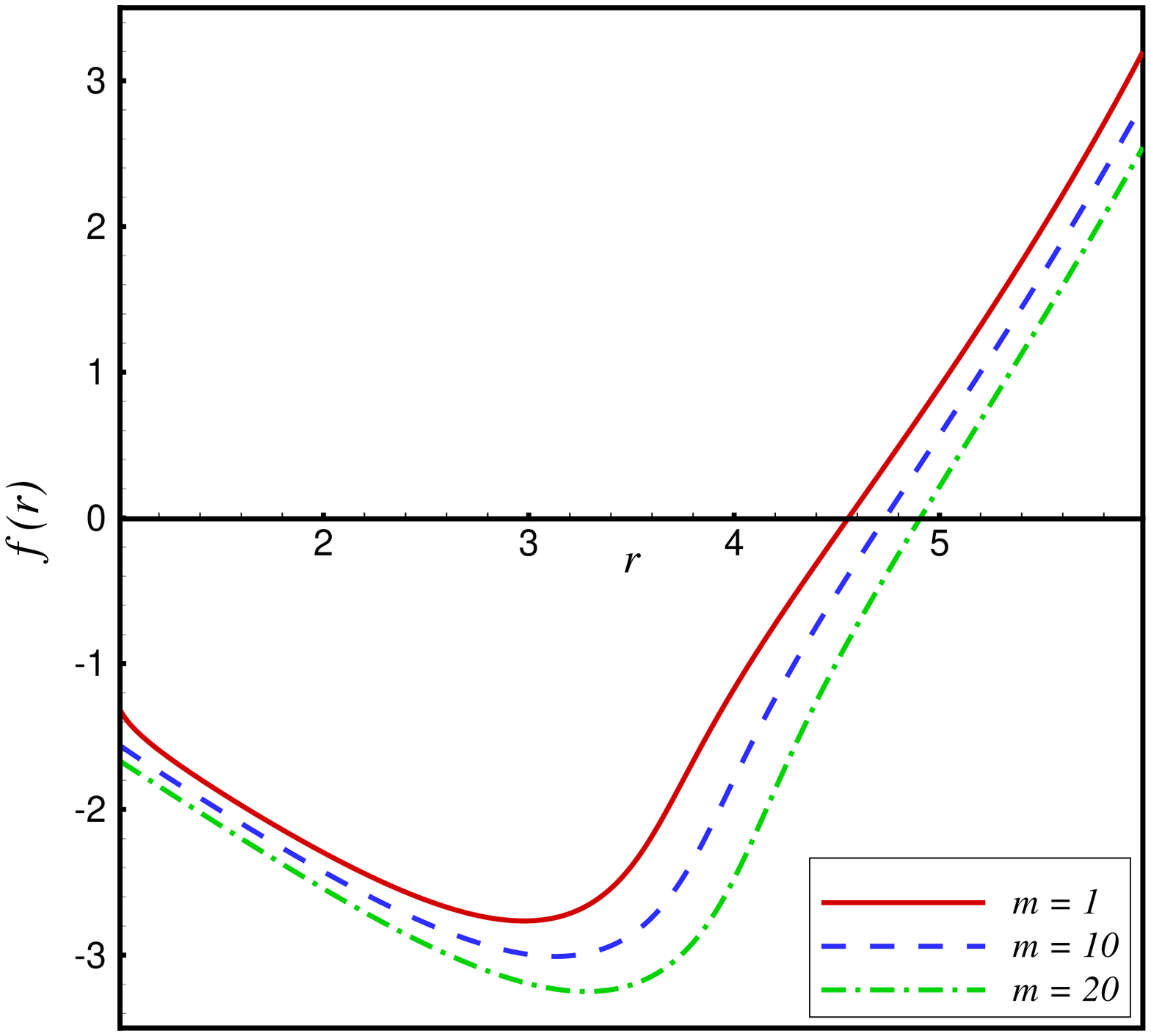}\label{Fig4a}}\hspace*{.2cm}
\subfigure[$f_{\mathrm{Einstein-Yang-Mills}}$]{\includegraphics[scale=0.27]{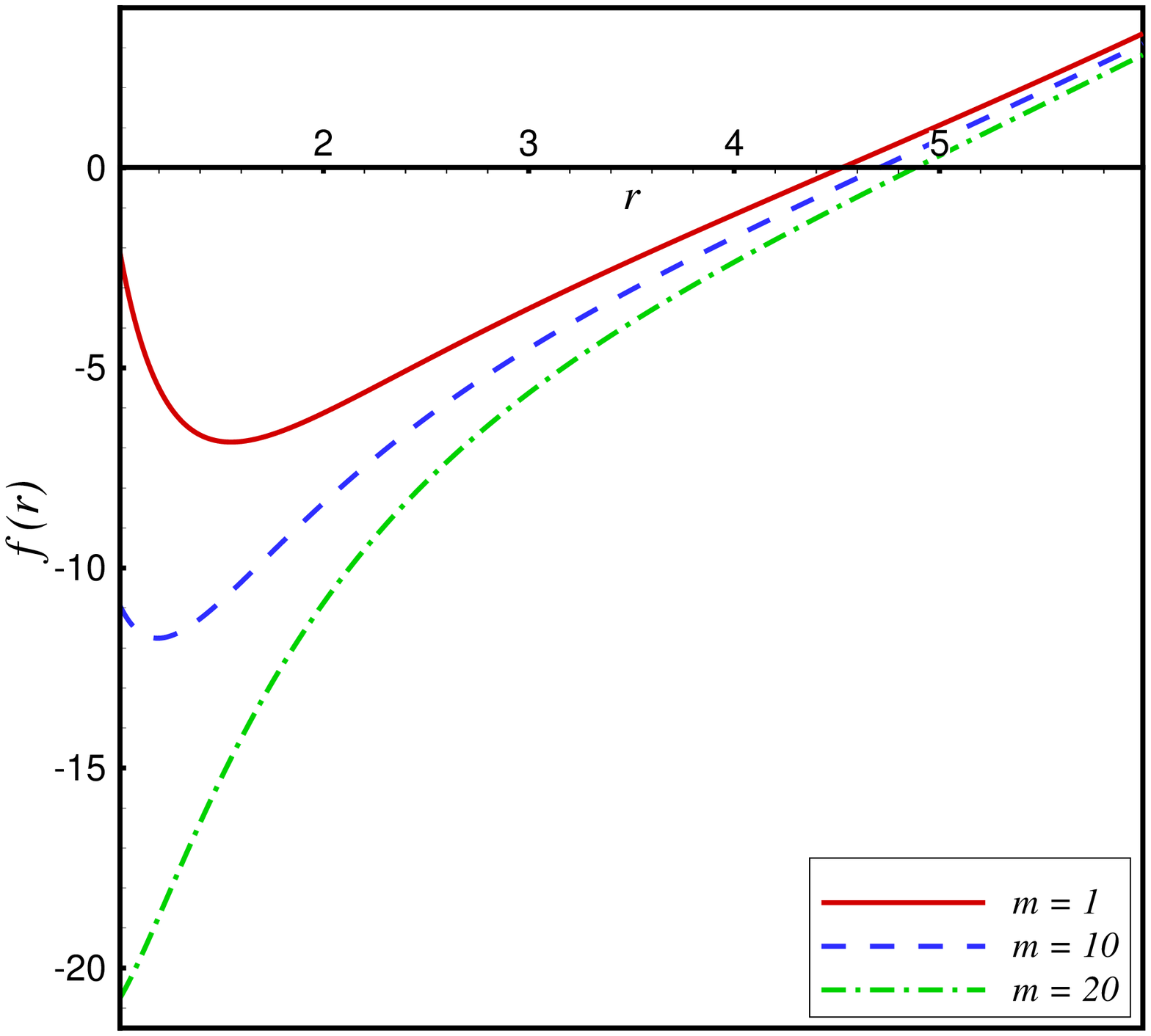}\label{Fig4b}}\caption{$f(r)$ with respect to $r$ for 5-dimensional spacetime($n=4$) and different values of the mass parameter $m$ with  $k=-1$, $e=4$ and $\Lambda=-1$.}\label{Fig4}
\end{figure}

Now, we probe the behavior of the quasitopological Yang-Mills black hole solution for the general case $\hat{\mu}_{i}\neq 0$, using Eq.\eqref{func11}. For this purpose, we have plotted $f(r)$ versus $r$ in Figs. \ref{Fig2}-\ref{Fig4} for $\hat{\mu}_{4}>0$ with $l=1$. These figures indicate that in the presence of the quasitopological gravity in low dimensions, the laps function goes to the constant value $k$, as $r\rightarrow 0$. In Fig.\ref{Fig2}, we have investigated the obtained six-dimensional($n=5$) solution for different values of $e$. Depending on the parameter $e$ in Fig.\ref{Fig2}, the solution can lead to a black hole with three horizons(solid red line and dashed blue line) or a naked singularity(dashed-dot green line). It should be noted that the largest horizon is the cosmological horizon and the smaller ones are the black hole horizons. \\
In Fig.\ref{Fig3}, we can observe the behavior of $f(r)$ with respect to $r$ for different values of the parameter $\hat{\mu}_{4}$ in space dimension, $n=6$. We can see that for the mentioned parameters $\hat{\mu}_{2}$, $\hat{\mu}_{3}$ $k$, $n$, $m$, $\Lambda$ and $\hat{\mu}_{4}$, there are black holes with inner and outer horizons, $r_{-}$ and $r_{+}$. The outer horizon is independent of the parameter $\hat{\mu}_{4}$, while the inner one increases as $\hat{\mu}_{4}$ increases. We can conclude that the behavior of the metric function is independent of the parmater $\hat{\mu}_{4}$ at the infinity. \\
We have also checked out the behavior of the metric function in the five-dimensional $(n=4)$ quasitopological gravity for different values of the mass parameter, $m$ in Fig.\ref{Fig4a} and then compared with the Einstein-Yang-Mills solution in Fig.\ref{Fig4b}. According to Eq.\ref{zeta1}, we should choose the region $r>1$ in order to have real solutions in five dimensions. Figure \ref{Fig4a} shows that there is a nonextreme black hole with a horizon $r_{+}$ for all values of the mass parameter $m$ with the mentioned parameters. As the parameter $m$ increases, the horizon $r_{+}$ also increases. Comparing these results with the Einstein-Yang-Mills solutions in Fig.\ref{Fig4b}, we find that the laps function has a large value at $r=1$. This is while the quasitopological Yang-Mills solution has a small value at this limit in Fig.\ref{Fig4a}.  \\
As the Kretchmann scalar diverges at $r=0$ for the positive finite value of $\mu_4$, so there is an essential singularity at the origin for these solutions.\\

\section{Thermodynamic behaviors of the quasitopological Yang-Mills black hole}\label{thermoY}
\begin{figure}
\centering
\subfigure[$e=1$]{\includegraphics[scale=0.27]{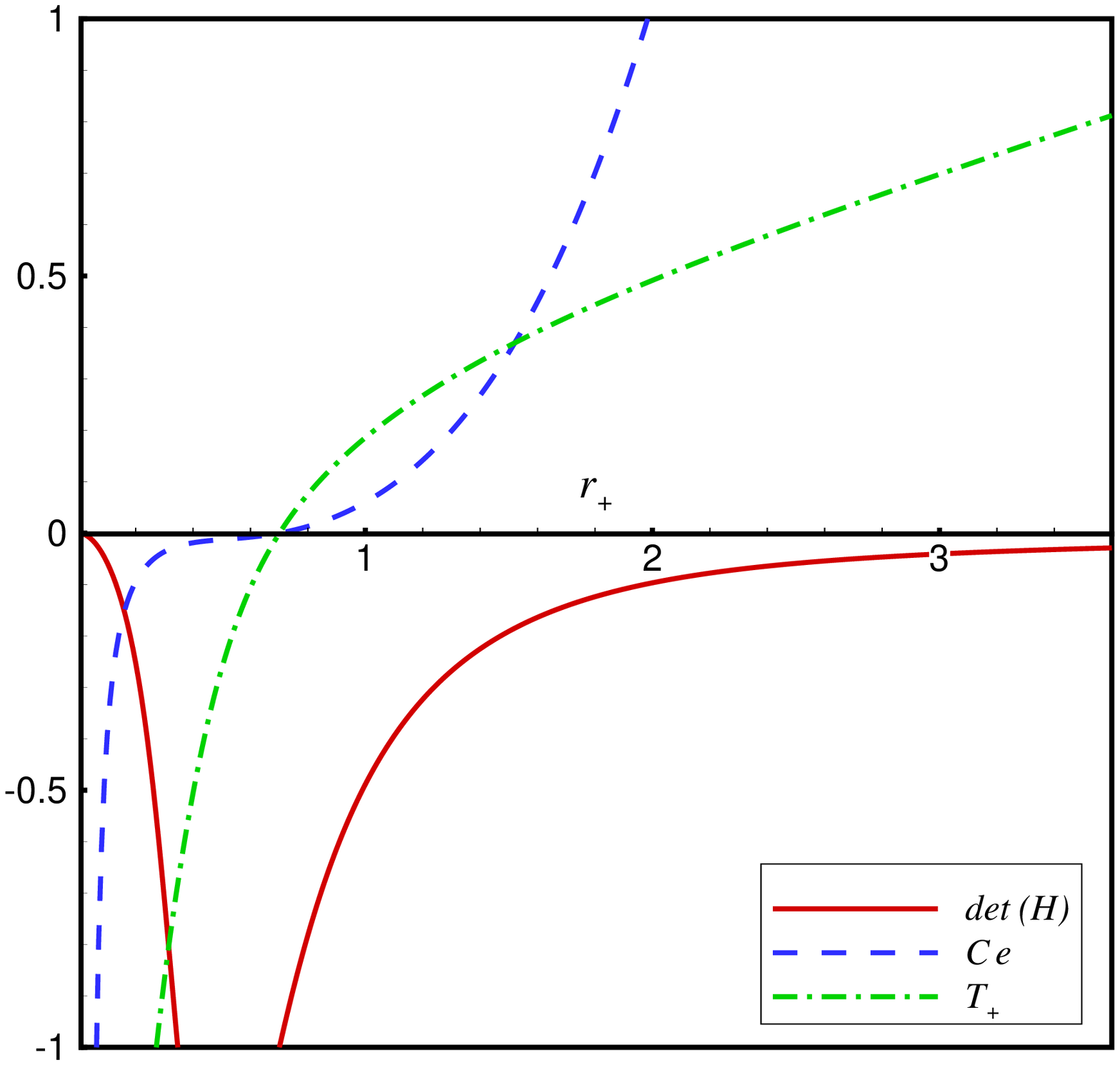}\label{Fig5a}}\hspace*{.2cm}\subfigure[$e=2$]{\includegraphics[scale=0.27]{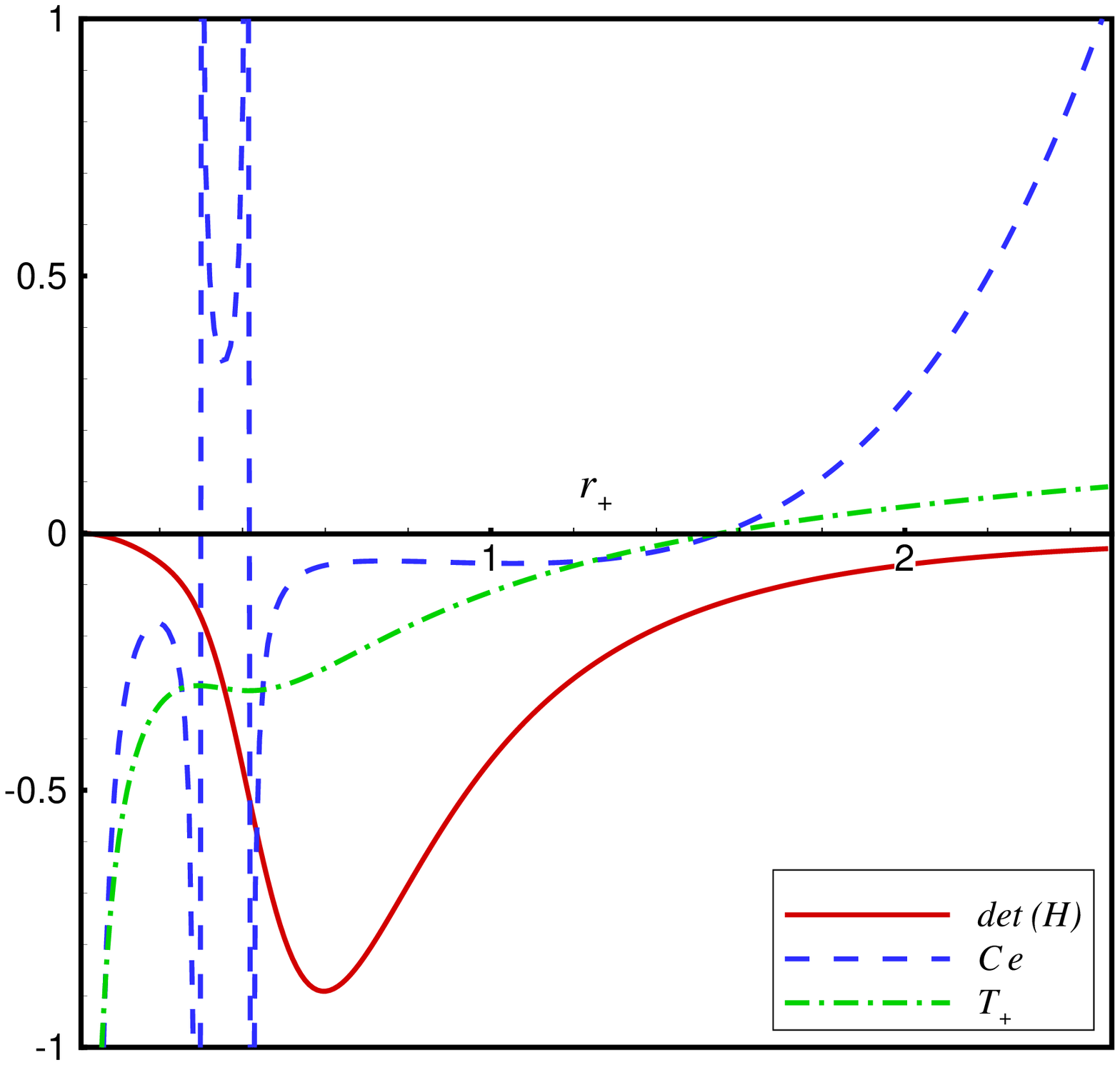}\label{Fig5b}}\caption{Thermal stability of Quasitopological Yang-Mills black hole with respect to $r_{+}$ with $\hat{\mu}_{2}=0.3$, $\hat{\mu}_{3}=0.1$, $\hat{\mu}_{4}=10^{-7}$, $k=1$, $n=5$ and $\Lambda=-1$.}\label{Fig5}
\end{figure}

\begin{figure}
\centering
\subfigure[$e=1$]{\includegraphics[scale=0.27]{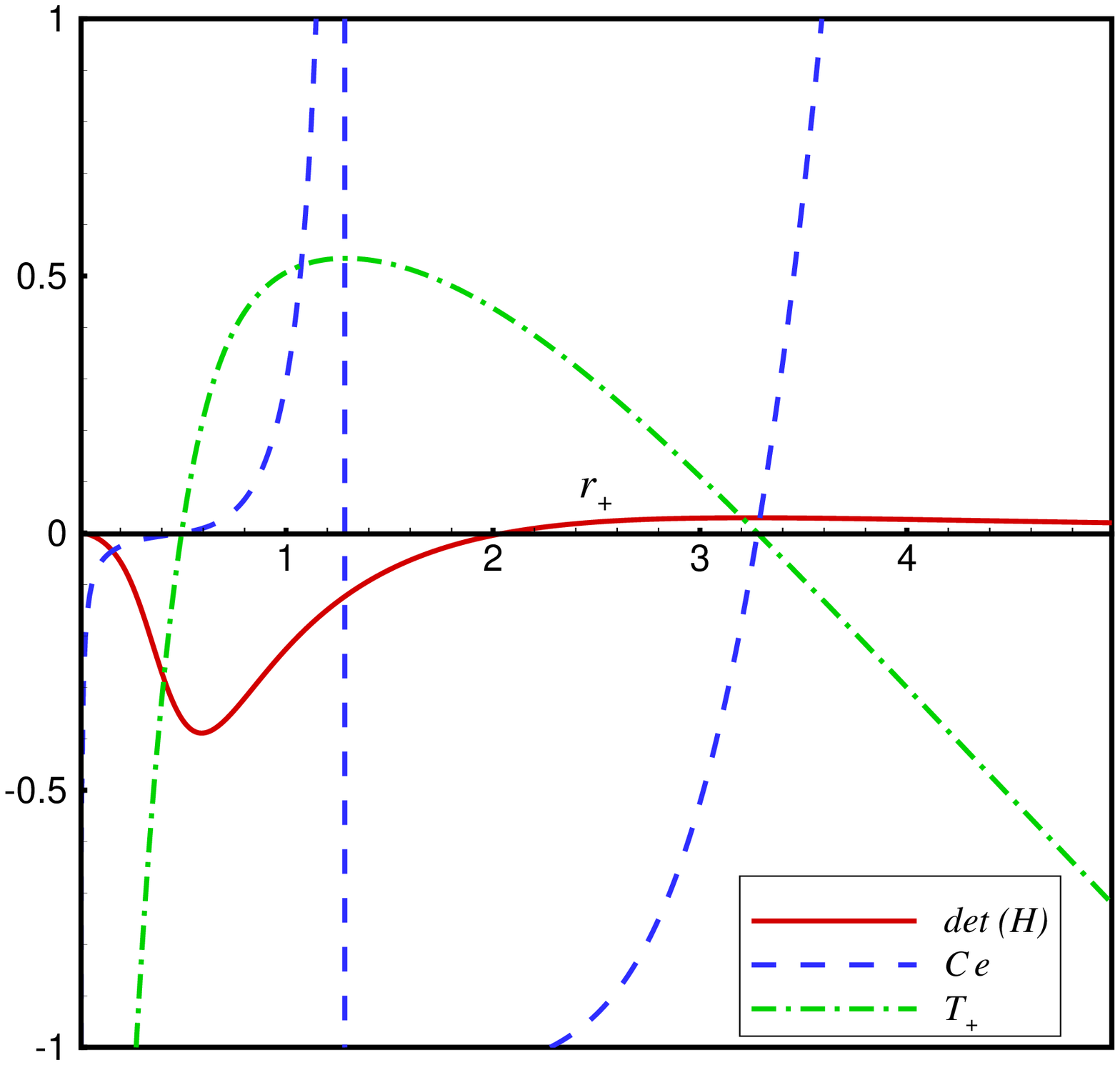}\label{Fig6a}}\hspace*{.2cm}\subfigure[$e=2$]{\includegraphics[scale=0.27]{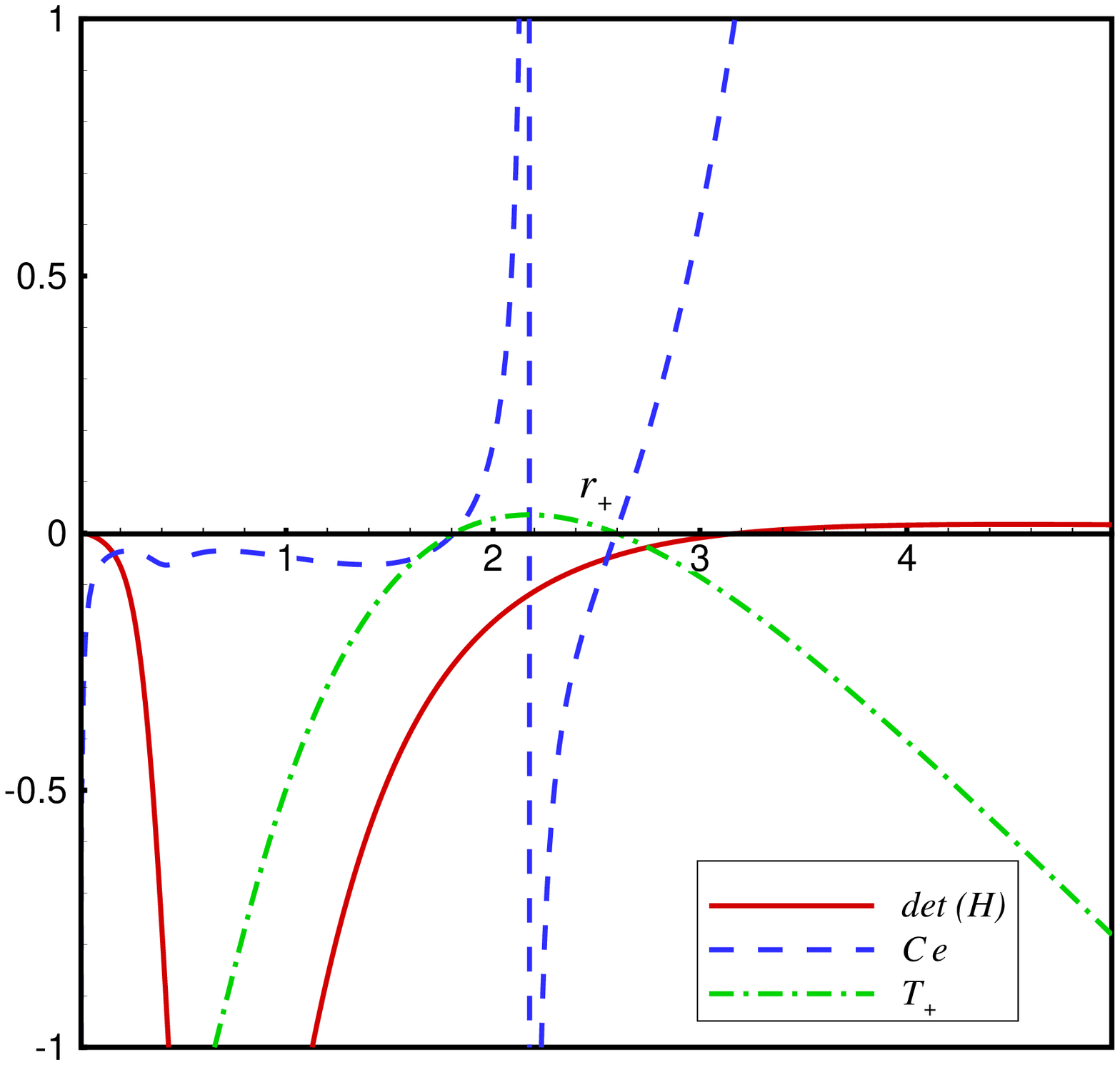}\label{Fig6b}}\caption{Thermal stability of Quasitopological Yang-Mills black hole with respect to $r_{+}$ with $\hat{\mu}_{2}=0.3$, $\hat{\mu}_{3}=0.2$, $\hat{\mu}_{4}=10^{-7}$, $k=1$, $n=6$ and $\Lambda=1$.}\label{Fig6}
\end{figure}

In this part, we want to calculate the thermodynamic quantities
such as mass, Yang-Mills charge, temperature, entropy and Yang-Mills potential of the quasitopological Yang-Mills black hole. We also probe the thermal stability of this black hole in both the canonical and the grand canonical ensembles. Using the subtraction method \cite{Brown}, we obtain the Arnowitt Deser Misner mass per unit volume $\omega_{n-1}$ (the volume of a (n-1)-dimensional unit sphere) of this black hole as follows
\begin{eqnarray}\label{massP}
M=\frac{(n-1)}{16\pi}m,
\end{eqnarray}
where the mass parameter $m$ is gained from the equation $f(r_{+})=0$,  
\begin{eqnarray}\label{maassY}
m(r_{+})&=&\left\{
\begin{array}{ll}
$$ \frac{\mu_{4}k^4}{r_{+}^{8-n}}+\frac{\mu_{3}k^3}{r_{+}^{6-n}}+\frac{\mu_{2}k^2}{r_{+}^{4-n}}+\frac{k}{r_{+}^{2-n}}-\frac{2\Lambda}{n(n-1)}r_{+}^n-\frac{(n-2)e^2}{(n-4)r_{+}^{4-n}}$$,\quad \quad \quad\quad\,\,\, \  \ {n>4,}\quad &  \\ \\
$$\frac{\mu_{4} k^4}{r_{+}^4}+\frac{\mu_{3} k^3}{r_{+}^2}+\mu_{2} k^2+kr_{+}^2-\frac{\Lambda}{6}r_{+}^{4}-2e^2\mathrm{ln}(r_{+})$$, \quad \quad\quad\quad\quad\quad\quad\quad{n=4.}\quad &
\end{array}
\right.
\end{eqnarray}
The Yang-Mills charge of this black hole per unit volume $\omega_{n-1}$ can be determined from the Gauss law
\begin{eqnarray}
Q=\frac{1}{4\pi\sqrt{(n-1)(n-2)}}\int d^{n-1}r \sqrt{\mathrm{Tr}(F_{\mu\nu}^{(a)}F_{\mu\nu}^{(a)})}=\frac{e}{4\pi}.
\end{eqnarray}
If we differentiate Eq. \eqref{EEE1} with respect to $r$ and use the fact that $f(r_{+})=0$, the Hawking temperature of the quasitopological Yang-Mills black hole can be derived from
\begin{eqnarray}\label{Tem2Y}
T_{+}&=&\frac{|f^{'}(r_{+})|}{4\pi}=\nonumber\\
&&\frac{|(n-8)\mu_{4}k^4+(n-6)\mu_{3}k^3r_{+}^2+(n-4)\mu_{2}k^2 r_{+}^4+k(n-2)r_{+}^6-\frac{2\Lambda r_{+}^8}{n-1}-(n-2)e^2 r_{+}^4|}{4\pi r_{+}|(4\mu_{4} k^3+3k^2\mu_{3}r_{+}^2+2k\mu_{2}r_{+}^4+r_{+}^6)|}.
\end{eqnarray}
Relation \eqref{Tem2Y} shows that the solutions with negative cosmological constant may have a larger range of parameters with positive temperature than the ones with positive cosmological constant.
Using Ref.\cite{wald}, we can obtain the entropy density of the quasitopological Yang-Mills black hole
\begin{eqnarray}\label{entropyY}
S&=&\frac{r_{+}^{n-1}}{4}+\frac{(n-1)k\mu_{2}}{2(n-3)}r_{+}^{n-3}+\frac{3(n-1)k^2\mu_{3}}{4(n-5)}r_{+}^{n-5}+\frac{(n-1)k^3\mu_{4}}{(n-7)}r_{+}^{n-7}.
\end{eqnarray}
Now, if we consider the mass $M$ as a function of $S$ and $Q$, the first law of the thermodynamics is established as 
\begin{eqnarray}\label{first}
dM=TdS+UdQ,
\end{eqnarray}
where $T=\big(\frac{\partial M}{\partial S}\big)_{Q}$ and $U=\big(\frac{\partial M}{\partial Q}\big)_{S}$. The calculations show that the evaluated $T$ is the same as Eq.\eqref{Tem2Y}. Using the relation $U=\big(\frac{\partial M}{\partial Q}\big)_{S}$, the Yang-Mills potential can be obtained as
\begin{eqnarray}\label{potenY}
U_{+}=\left\{
\begin{array}{ll}
$$-\frac{2\pi Q (n-1)(n-2)}{(n-4)}r_{+}^{n-4}$$,\quad\quad\quad\quad\quad  \ {n>4}\quad &  \\ \\
$$-4\pi Q (n-1)\mathrm{ln}(r_{+})$$\,\quad\quad\quad\quad\quad\,\,\,\,  \ {n=4.}\quad &  \\ \\
\end{array}
\right.
\end{eqnarray}
We can also probe the thermal stability of the quasitopological Yang-Mills black hole in both the canonical and the grand canonical ensembles. In the canonical ensemble, the electric charge $Q$ is fixed, and so, we analyze the stability of the black hole considering the small variation of the entropy $S$. Thereby, thermal stability in the canonical ensemble will be established if the heat capacity
\begin{eqnarray}
C_{e}=T\bigg(\frac{\partial S}{\partial T}\bigg)_{Q}=T\bigg(\frac{\partial^2 M}{\partial S^2}\bigg)^{-1}_{Q},
\end{eqnarray}
is positive. We should note that the positive value of the temperature is also a necessary condition in order to have a physical solution. In the grand canonical ensemble, both parameters $S$ and $Q$ are variables. For this ensemble, in addition to $T_{+}$ and $C_{e}$, the positive values of the parameters $\Big(\frac{\partial ^2 M}{\partial Q^2}\Big)$ and the Hessian matrix determinant  
\begin{eqnarray}
detH=\Big(\frac{\partial ^2 M}{\partial S^2}\Big)\Big(\frac{\partial ^2 M}{\partial Q^2}\Big)-\Big(\frac{\partial ^2 M}{\partial S\partial Q}\Big)^2,
\end{eqnarray}
guarantee the thermal stability of the black hole. If we calculate $\Big(\frac{\partial ^2 M}{\partial Q^2}\Big)$ for this black hole, we get to
\begin{eqnarray}
\Big(\frac{\partial ^2 M}{\partial Q^2}\Big)=\left\{
\begin{array}{ll}
$$-\frac{2\pi (n-1)(n-2)}{(n-4)}r_{+}^{n-4}$$,\quad\quad\quad\quad\quad  \ {n>4,}\quad &  \\ \\
$$-4\pi (n-1)\mathrm{ln}(r_{+})$$\,\quad\quad\quad\quad\quad\,\,\,\,  \ {n=4,}\quad &  \\ \\
\end{array}
\right.
\end{eqnarray}
which is negative for $n>3$. As the quasitopological Yang-Mills black hole does not satisfy one of the conditions of thermal stability in the grand canonical ensemble, so this black hole is not thermally stable in this ensemble. In order to recognize the stability regions of the quasitopological Yang-Mills black hole, we have plotted figures \ref{Fig5} and \ref{Fig6} for $\Lambda<0$ and $\Lambda>0$. In Fig. \ref{Fig5a} with $\Lambda<0$, there is a $r_{+\mathrm{QY}}$ which both $T_{+}$ and $C_{e}$ in the six-dimensional quasitopological Yang-Mills black hole are positive for $r_{+}>r_{+\mathrm{QY}}$ and thus this black hole is stable in this range for the mentioned parameters in the caption. As the value of the Yang-Mills charge $e$ increases in Fig.\ref{Fig5b}, the value of $r_{+\mathrm{QY}}$ increases. Therefore we can conclude that for the mentioned parameters, the solutions may have a larger region in thermal stability for the smaller values of $e$. According to our above statement, detH is not positive for the mentioned region $r_{+}>r_{+\mathrm{QY}}$ and thus this black hole is not stable in the grand canonical ensemble. In Fig.\ref{Fig6}, we have depicted the stability of the quasitopological Yang-Mills black hole for $\Lambda>0$. Figure \ref{Fig6} manifests that in the dS spacetime like the AdS one, the quasitopological Yang-Mills solutions have a larger range of parameters in thermal stability for the small values of $e$ than the large ones. Comparing figures \ref{Fig5} and \ref{Fig6} also shows that the solutions with $\Lambda>0$ have a smaller stable region in comparison to the ones with $\Lambda<0$.
\section{Critical behavior of the quasitopological Yang-Mills black holes in the extended phase space}\label{critical1}
Until now, a lot of studies about the black hole phase transition have been done where some new researches are in Refs. \cite{Mirza1,Mirza2,Mirza3,Mirza4}. In this section, we investigate the critical behavior and phase transition of the quasitopological Yang-Mills black hole. We consider an extended phase space in which the quantities $S$, $Q$, the cosmological constant $\Lambda$, the Gauss-Bonnet, the cubic and quartic quasitopological coefficients ($\hat{\mu}_{i}$) and their conjugates are considered as the thermodynamic variables. We consider $\Lambda$ as a thermodynamic pressure by the relation $P=-\frac{\Lambda}{8\pi}$, which its conjugate quantity( the thermodynamic volume) is defined by the relation $V=\frac{r_{+}^n}{n}$ \cite{Dolan}. We also introduce the specific volume $v=\frac{4r_{+}}{n-1}$ to compare the equation of state with the Van der Walls equation. By these definitions, the first law of the black hole thermodynamics in the extended phase space can be written \cite{Henn} as follows
\begin{eqnarray}
dM=TdS+UdQ+VdP+\Psi_{2}d\hat{\mu}_{2}+\Psi_{3}d\hat{\mu}_{3}+\Psi_{4}d\hat{\mu}_{4},
\end{eqnarray}
where $\Psi_{i}$ is denoted as the conjugate of the coefficient $\hat{\mu}_{i}$ which can be obtained from Eqs.\eqref{massP} and \eqref{maassY} as follows 
\begin{eqnarray}
\Psi_{2}&=&\frac{\partial M}{\partial \hat{\mu}_{2}}=\frac{k^2(n-1)}{16\pi }r_{+}^{n-4}-\frac{k(n-1)}{2(n-3) }r_{+}^{n-3}T_{+},\nonumber\\
\Psi_{3}&=&\frac{\partial M}{\partial \hat{\mu}_{3}}=\frac{k^3(n-1)}{16\pi }r_{+}^{n-6}-\frac{3k^2(n-1)}{4(n-5) }r_{+}^{n-5}T_{+},\nonumber\\
\Psi_{4}&=&\frac{\partial M}{\partial \hat{\mu}_{4}}=\frac{k^4(n-1)}{16\pi }r_{+}^{n-8}-\frac{k^3(n-1)}{(n-7) }r_{+}^{n-7}T_{+}.
\end{eqnarray}
To study the critical behavior of the quasitopological Yang-Mills black hole, we first plot the $P-v$ isotherms and the $G-T$ diagrams and then obtain the critical exponents in the following subsections:
\subsection{$P-v$ isotherm} 
\begin{figure}
\center
\includegraphics[scale=0.5]{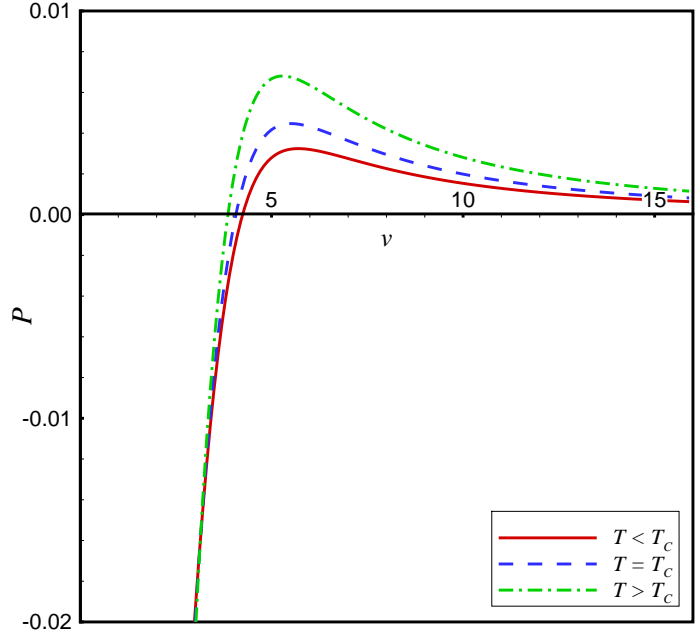}
\caption{\small{P versus $v$ for different temperature $T$ with $\hat{\mu}_{2}=0.2$, $\hat{\mu}_{3}=0.1$ $\hat{\mu}_{4}=0.001$, $k=1$ and $n=4$ $e=0.2$.} \label{Fig7}}
\end{figure}

$P-v$ isotherm is one of the candidates by which we can compare the critical behavior of a black hole with the Van der Walls fluid. In this section, we probe $P-v$ isotherm for the obtained black hole. From the relation $\Lambda=-8\pi P$ and Eq.\eqref{Tem2Y}, the equation of state can be found as
\begin{eqnarray}\label{pres}
P&=&\frac{T}{v}-\frac{k(n-2)}{(n-1)\pi v^2}+\frac{16(n-2)e^2}{(n-1)^3\pi v^4}+\frac{32k\mu_{2}}{ (n-1)^2v^3}\bigg(T-\frac{k(n-4)}{2\pi(n-1) v}\bigg)+\frac{768k^2\mu_{3}}{ (n-1)^4v^5}\bigg(T-\frac{k(n-6)}{3\pi(n-1) v}\bigg)\nonumber\\
&&+\frac{16384k^3\mu_{4}}{ (n-1)^6v^7}\bigg(T-\frac{k(n-8)}{4\pi(n-1) v}\bigg).
\end{eqnarray}
The critical points can be derived from the following conditions: 
\begin{eqnarray}\label{cons}
\frac{\partial P}{\partial v}=0\,\,\,,\,\,\,\frac{\partial ^2 P}{\partial v^2}=0.
\end{eqnarray}
We specify the volume, pressure and temperature of the critical points such as $v_{C}$, $P_{C}$ and $T_{C}$.
By applying the above conditions \eqref{cons} to the equation of state \eqref{pres}, we can not find any analytic solutions for the critical points and therefore, we use a numeric method. The results show that there is no critical behavior for the hyperbolic case, $k=-1$ and therefore, we concentrate on the spherical case with $k=1$. By considering three cases $T<T_{C}$, $T=T_{C}$ and $T>T_{C}$, we have plotted $P$ vs $v$ for $\hat{\mu}_4>0$ and $k=1$ in Fig.\ref{Fig7}. This figure shows that the critical behavior of the quasitoplogical Yang-Mills black hole is similar to that of the Van der Walls fluid. 

\subsection{$G-T$ diagrams}
\begin{figure}
\center
\includegraphics[scale=0.5]{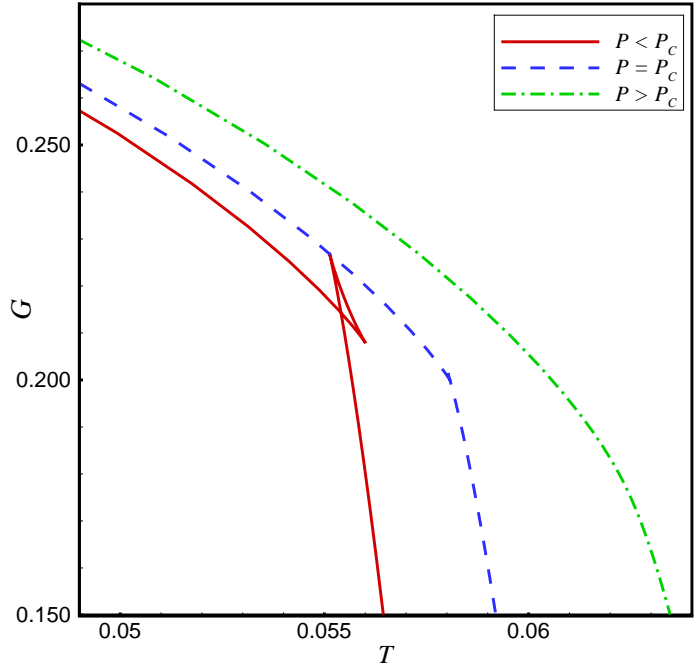}
\caption{\small{G versus $T$ for different pressure $P$ with $\hat{\mu}_{2}=0.2$, $\hat{\mu}_{3}=0.1$ $\hat{\mu}_{4}=0.001$, $k=1$ and $n=4$ $e=0.2$.} \label{Fig8}}
\end{figure}

Another candidate for probing the critical behavior of a black hole is the $G-T$ diagram(Gibbs free energy $G$ versus temperature $T$). The Gibbs energy for the quasitopological Yang-Mills black hole is gained by 
\begin{eqnarray}
G=M-TS.
\end{eqnarray}
We have plotted $G$ vs $T$ in Fig.\ref{Fig8}. It shows that for $P<P_C$, there is a swallowtail behavior which informs of a first-order phase transition from a small to a large black hole in the quasitopological Yang-Mills theory.

\subsection{critical exponents}
To describe the physical behavior near the critical points, we aim to obtain the critical exponents $\alpha$, $\beta$, $\gamma$ and $\delta$ for the quasitopological Yang-Mills black hole. As the entropy of this black hole is independent of the temperature $T$ in Eq.\eqref{entropyY}, we conclude that  
\begin{eqnarray}
C_{V}=T\bigg(\frac{\partial S}{\partial T}\bigg)_{V}=0\,\,\Rightarrow\,\,\,\alpha=0.
\end{eqnarray} 
Let us substitute the following definitions 
\begin{eqnarray}
p=\frac{P}{P_{C}}\,\,\,\,\,,\,\,\,\,\,\nu=\frac{v}{v_{C}}\,\,\,\,\,,\,\,\,\,\,\tau=\frac{T}{T_{C}},
\end{eqnarray}
in Eq.\eqref{pres} and expand it near the critical point, $\tau=1+t$ and $\nu=(1+\omega)^{1/z}$. So we obtain 
\begin{eqnarray}\label{pa}
p=1+\frac{t}{\rho_{C}}-\frac{1}{z\rho_{C}}tw-Aw^3+\mathcal{O}(tw^2,w ^4),
\end{eqnarray}
where
\begin{eqnarray}
A&=&\frac{1}{z^3}\bigg(\frac{1}{\rho_{c}}-\frac{h^{(3)}|_{\nu=1}}{6}\bigg)\,\,\,\,\,\,,\,\,\,\,\,\,\rho_{C}=\frac{P_{C}v_{C}}{T_{C}}\,\,\,\,\,\,,\,\,\,\,\,\,\nonumber\\
h(\nu)&=&\frac{1}{P_{C}}\bigg[-\frac{k(n-2)}{(n-1)\pi \nu^2v_C^2}+\frac{16(n-2)e^2}{(n-1)^3\pi \nu^4v_C^4}+\frac{32k\mu_{2}}{ (n-1)^2\nu^3v_C^3}\bigg(\tau T_{C}-\frac{k(n-4)}{2\pi(n-1) \nu v_C}\bigg)\nonumber\\
&&+\frac{768k^2\mu_{3}}{ (n-1)^4\nu^5v_C^5}\bigg(\tau T_{C}-\frac{k(n-6)}{3\pi(n-1) \nu v_C}\bigg)+\frac{16384k^3\mu_{4}}{ (n-1)^6\nu^7v_C^7}\bigg(\tau T_{C}-\frac{k(n-8)}{4\pi(n-1) \nu v_C}\bigg)\bigg].
\end{eqnarray} 
Differentiating Eq.\eqref{pa} with respect to $w$ and imposing the Maxwell's equal area law, we get to
\begin{eqnarray}\label{aa}
p&=&1+\frac{t}{\rho_{C}}-\frac{1}{z\rho_{C}}tw_{l}-Aw_{l}^3=1+\frac{t}{\rho_{C}}-\frac{1}{z\rho_{C}}tw_{s}-Aw_{s}^3,\nonumber\\
0&=&\int_{\omega_{l}}^{\omega_{s}}\omega dP,
\end{eqnarray}
where $w_{l}$ and $w_{s}$ are denoted as the ``volume" of large and small black holes. Eq.\eqref{aa} has the following unique solution
\begin{eqnarray}
w_{s}=-\omega_{l}=\sqrt{-\frac{t}{z \rho_{C} A}},
\end{eqnarray}
which yields to
\begin{eqnarray}
\eta=v_{C}(\omega_{l}-\omega_{s})=2v_{C}\omega_{l}\propto\sqrt{-t}\,\,\Rightarrow\,\,\beta=\frac{1}{2}.
\end{eqnarray}
The isothermal compressibility can also be obtained from Eq.\eqref{pa} as
\begin{eqnarray}
\kappa_{T}=-\frac{1}{V}\frac{\partial V}{\partial P}|_{T}\propto\frac{\rho_{C}}{P_{C}}\frac{1}{t}\,\,\Rightarrow\,\,\,\gamma=1.
\end{eqnarray}
Finally, we can obtain the ``shape" of the critical isothem $t=0$, from Eq.\eqref{pa}
\begin{eqnarray}
p-1=-C w^3\,\,\Rightarrow\,\,\,\delta=3.
\end{eqnarray} 
The obtained results show that the critical exponents of the quasitopological Yang-Mills black hole coincide
with the ones for the Van der Waals fluid.\\
In the first part of this paper, we obtained the quasitopological Yang-Mills black hole solutions and then investigated their physical and thermodynamic behaviors. In the second part, we are willing to obtain a new class of pure quasitopological Yang-Mills black hole solutions and probe their behaviors. 

\section{pure quasitopological Yang-Mills black hole solutions}\label{pure}
\begin{figure}
\centering
\includegraphics[scale=0.5]{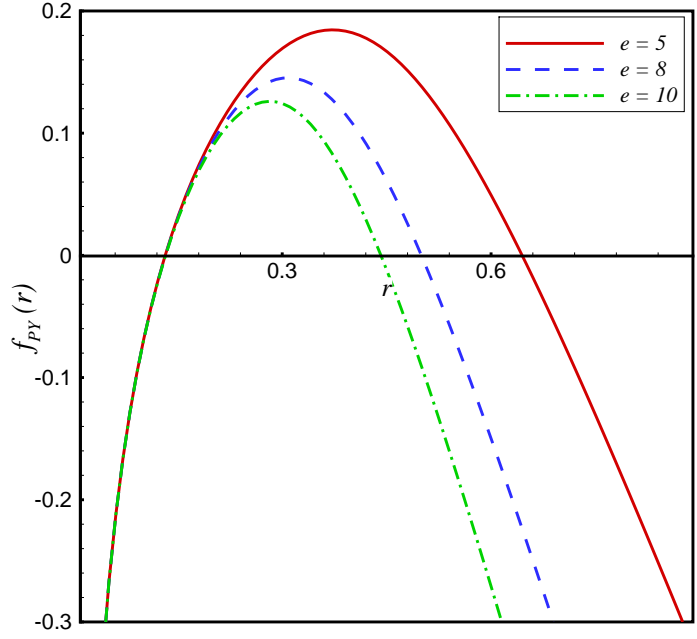}
\caption{\small{$f_{PY}(r)$ with respect to $r$ for different values of $e$ with $\hat{\mu}_{4}=10^{-7}$, $k=1$, $n=9$, $m=1$ and $\Lambda=1$. There are two horizons which the left one is the black hole horizon and the right one is the cosmological horizon.} \label{Fig9}}
\end{figure}

\begin{figure}
\centering
\includegraphics[scale=0.5]{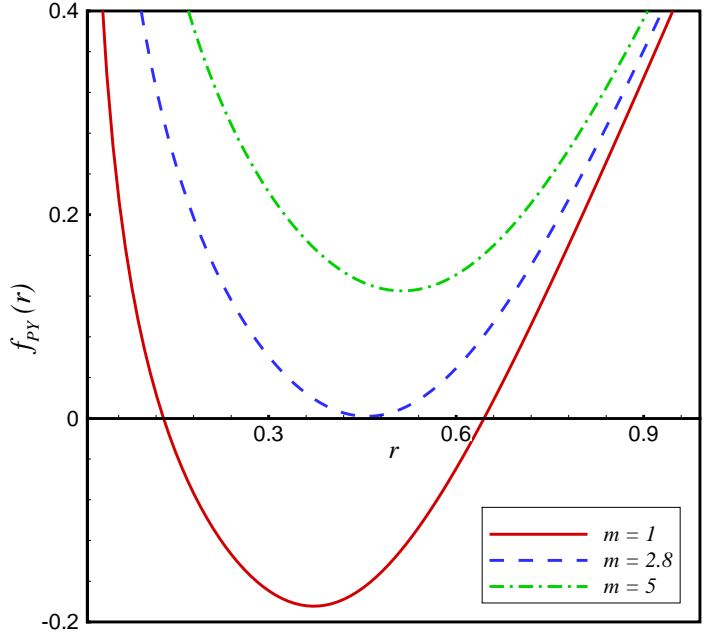}
\caption{\small{$f_{PY}(r)$ with respect to $r$ for different values of $m$ with $\hat{\mu}_{4}=10^{-7}$, $k=-1$, $e=5$, $n=9$ and $\Lambda=1$.} \label{Fig10}}
\end{figure}

In this section, we first define the pure quasitopological action with Yang-Mills theory and then obtain the related black hole solutions. For this purpose, we choose ${\mathcal L}_1={\mathcal L}_2={\mathcal L}_3=0$, where it leads to the action 
\begin{equation}\label{Act2}
I_{\rm{bulk}}=\frac{1}{16\pi}\int{d^{n+1}x\sqrt{-g}\big\{-2\Lambda+\hat{\mu}_{4}\mathcal{L}_{4}-\gamma_{ab}F_{\mu\nu}^{(a)}F^{(b)\mu\nu}\big\}}.
\end{equation}
Therefore, the gravitational field equation \eqref{EEE1} gets to 
\begin{eqnarray}\label{EE1}
\mu_{4} \Psi^4+\zeta=0,
\end{eqnarray}
where we have defined $\Psi$ and $\zeta$ in Eq. \eqref{zeta1}. For this equation, we arrive at the solution
\begin{eqnarray}\label{Eqq}
f_{PY}(r)&=&k\mp\frac{r^{\frac{3}{2}}}{\mu_{4}}\times\left\{
\begin{array}{ll}
$$\bigg[\mu_{4}^3\big(\frac{2\Lambda r^2}{n(n-1)}+\frac{(n-2)e^2}{(n-4)r^2}+\frac{m}{r^{n-2}}\big)\bigg]^{\frac{1}{4}}$$,\quad\quad\,\  \ {n>4,}\quad &  \\ \\
$$\bigg[\mu_{4}^3\big(\frac{\Lambda r^2}{6}+\frac{2e^2\mathrm{ln}(\frac{r}{r_{0}})}{r^2}+\frac{m}{r^{2}}\big)\bigg]^{\frac{1}{4}}$$, \quad\quad\quad\quad\quad{n=4,}\quad &
\end{array}
\right.
\end{eqnarray}
where we choose $r_{0}=1$. In order to have real solutions, we should consider a positive value for $\Lambda$ and $\mu_{4}$. Since all dimensions except $n=8$ lead to a positive value for $\mu_{4}$, so we ignore this dimension. Our numerical results also show that in order to have a black hole, the spacetime dimension should be larger than eight.  The pure quasitopological Yang-Mills black hole has a horizon, if the equation $f_{PY}(r_{+})=0$ has a real solution. So, depending on the values of the parameters $n$, $m$, $e$, $\Lambda$ and $\mu_{4}$, the solutions may lead to a black hole with horizons. For $r\rightarrow\infty$, $f_{PY}(r)$ goes to 
\begin{eqnarray}\label{inf}
f_{PY}(r)=k\mp\big(\frac{2\Lambda}{\mu_{4} n(n-1)}\big)^{\frac{1}{4}}r^2,
\end{eqnarray}
which shows that we should define $"-"$ and $"+"$ for repsectively $k=+1$ and $k=-1$, otherwise we face a naked singularity. So, we can conclude that with $\Lambda>0$ in Eq.\eqref{inf}, the solutions in pure quasitopological Yang-Mills gravity may lead to the asymptotically AdS and dS black holes with $k=-1$ and $k=+1$, respectively.\\
For better understanding, we have plotted $f_{PY}(r)$ versus $r$ with positive cosmological constant in Figs.\ref{Fig9}-\ref{Fig10}. In Fig.\ref{Fig9}, there are two small and large horiozons for all black hole which are respectively related to the balck hole and cosmological horizons.\\ 
In Fig. \ref{Fig10}, we have plotted AdS solutions with $k=-1$ and $n=9$ for different values of the mass parameters $m$. It shows that although the two above mentioned conditions (spacetime dimensions larger than eight and $k=-1$ for AdS solutions) are established, however depending on the parameters $e$, $n$, $\hat{\mu}_{4}$ and $m$, there is an AdS black hole with two horizons, an extreme dS black hole and a naked singularity. For the mentioned parameters, the solutions with small mass parameter $m$ may lead to a black hole with two horizons. \\
For a finite positive value of $\mu_4$ in the pure quasitopological Yang-Mills black hole solutions, the Kretschmann scalar diverges at the origin $r=0$ 
\begin{eqnarray}
R{abcd}R^{abcd}\propto\sqrt{\frac{m}{\mu_{4}}}r^{-\frac{n}{2}}.
\end{eqnarray}
Therefore, the pure quasitopological Yang-Mills black hole has an essensial singularity at $r=0$.

\section{Thermodynamic behaviors of the pure quasitopological Yang-Mills black holes}\label{thermoP}
We also search for the thermodynamic behaviors of the pure quasitopological Yang-Mills black holes. The mass of this black hole is also followed from Eq. \eqref{massP}, where the mass parameter is gained as  
\begin{eqnarray}\label{maassP}
m(r_{+})&=&\left\{
\begin{array}{ll}
$$ -\frac{2\Lambda}{n(n-1)}r_{+}^n-\frac{(n-2)e^2}{(n-4)r_{+}^{4-n}}+\frac{\mu_{4}k^4}{r_{+}^{8-n}}$$,\quad \quad \quad\quad\,\,\, \  \ {n>4,}\quad &  \\ \\
$$-\frac{\Lambda}{6}r_{+}^{4}-2e^2\mathrm{ln}(r_{+})+\frac{\mu_{4} k^4}{r_{+}^4}$$, \quad \quad\quad\quad\quad\quad\quad{n=4.}\quad &
\end{array}
\right.
\end{eqnarray}
Using Eq. \eqref{EE1} and the condition $f_{PY}(r_{+})=0$, the Hawking temperature and the entropy density of the pure quasitopological Yang-Mills black hole are followed from
\begin{eqnarray}\label{Tem2P}
T_{+}=\frac{|f^{'}(r_{+})|}{4\pi}=|\frac{k(n-8)}{16\pi r_{+}}-\frac{\Lambda r_{+}^7}{8\pi(n-1)\mu_{4}k^3}-\frac{(n-2)e^2r_{+}^3}{16\pi\mu_{4}k^3}|,
\end{eqnarray}
and 
\begin{eqnarray}\label{entropyP}
S&=&\frac{(n-1)\mu_{4}k^3}{(n-7)}r_{+}^{n-7}.
\end{eqnarray}
The pure quasitopological Yang-Mills black hole solutions obey from the first law of black hole thermodynamics if the Yang-Mills potential is followed from
\begin{eqnarray}\label{potenP}
U_{+}=\left\{
\begin{array}{ll}
$$-\frac{\Lambda r_{+}^n}{2n}-\frac{(n-1)(n-2)e^2}{4(n-4)r_{+}^{4-n}}+\frac{(n-1)\mu_{4} k^4}{4r_{+}^{8-n}}$$,\quad\quad\quad\quad\quad  \ {n>4}\quad &  \\ \\
$$-\frac{\Lambda r_{+}^4}{8}-\frac{3}{2}e^2\mathrm{ln}(r_{+})+\frac{3\mu_{4} k^4}{4r_{+}^4}$$\quad\quad\quad\quad\quad\quad\quad\quad  \ {n=4.}\quad &  \\ \\
\end{array}
\right.
\end{eqnarray}
To probe the thermal stability of the pure quasitopological Yang-Mills black hole, we calculate the heat capacity 
\begin{eqnarray}
C_{Q}=-\frac{16\pi \mu_{4}^2k^6(n-1)^2r_{+}^{n-6}}{14\Lambda r_{+}^8+3(n-1)(n-2)e^2r_{+}^4+(n-1)(n-8)\mu_{4}k^4}T_{+}.
\end{eqnarray}
This relation shows that for $\Lambda$ and $\mu_{4}>0$, it is not possible to have positive values for both $C_{Q}$ and $T_{+}$ simultaneously. So, the pure quasitopological Yang-Mills black hole is not thermally stable in the canonical and the grand canonical ensembles.\\

\section{concluding remarks}\label{con}
In this paper, we achieved two new sets of $(n+1)$-dimensional black hole solutions in the quasitopological gravity with the nonabelian Yang-Mills theory. Followed by AdS/CFT correspondence, the obtained solutions may provide vast backgrounds to study the $n$-dimensional CFT's. At first, we considered the quasitopological gravity up to the fourth-order curvature tensor and an $N$-parameters gauge group $\mathcal{G}$ with the structure constants $C^{i}_{jk}$. Using the Wu-Yang ansatz, we defined the gauge potentials which have the Lie algebra of the $SO(n)$ and $SO(n-1,1)$ gauge groups. For this theory, we obtained two types of the analytic quasitopological Yang-Mills black hole solutions for $\mu_{4}>0$ and $\mu_{4}<0$. Real solutions in the range $0<r<\infty$(which is described for $n>5$ and the range $1<r<\infty$ which is defined for $n=4$) were obtained only for $\mu_{4}>0$. We also probed the physical structures of the quasitopological Yang-Mills solutions in two cases. The first one is $\hat{\mu}_{2}=\hat{\mu}_{3}=0$, $\hat{\mu}_{4}\neq0$ and the other one is $\hat{\mu}_{i}\neq0$ ($i=2,3,4$). We proved that for small values of $\hat{\mu}_{i}$ ($i=2,3,4$), the quasitopological Yang-Mills solutions reduce to the Einstein-Yang-Mills ones plus some correction terms proportional to $\mu_{4}$. We also showed that unlike the Einstein's theory, the quasitopological gravity has the ability to provide a finite value for the metric function at the origin for $n\leq 8$. For the limit $r\rightarrow\infty$, the quasitopological gravity effect will be negligible and so the black hole has a similar behavior as the one in Einstein gravity. Depending on the values of the parameters $\hat{\mu}_{2}$, $\hat{\mu}_{3}$, $\hat{\mu}_{4}$, $e$, $m$, $\Lambda$ and $k$, we encountered a black hole with inner and outer horizons, an extreme black hole or a naked singularity. For $n\leq 8$, the Kreshmann scalar diverges at $r\rightarrow 0$ which we can deduce that there is an essential singularity at this point.\\
An investigation of the thermodynamic behaviors of the quasitopological Yang-Mills black hole was raised in this paper. We also checked out the accuracy of the first law of the thermodynamics and probed the thermal stability of this black hole in the both canonical and the grand canonical ensembles. The results showed that the quasitopological Yang-Mills black hole is thermally stable in just the canonical ensemble. We deduced that the solutions with negative cosmological constant and small Yang-Mills charge $e$ may lead to a larger range of parameters in thermal stability compared to the ones with the positive cosmological constant and large $e$.\\
We also probed the critical behavior of the quasitopological Yang-Mills black hole in the extended phase space in which the entropy $S$, the Yang-Mills charge $e$, the cosmological constant $\Lambda$, the coupling constants $\hat{\mu}_{i}$'s ($i=2,3,4$) and also their conjugate quantities are considered as the thermodynamic variables. We concluded that the $P-v$ isotherms with $k=1$ and the critical exponents of this black hole behave like the ones in the Van der Walls fluid. We also found a swallowtail behavior for the Gibbs free energy which showed a first order small-large black hole transition.  \\
In the second step, we obtained the pure quasitopological Yang-Mills black hole solutions. For this purpose, we just considered the cosmological constant, Yang-Mills and quartic quasitopological terms. Real pure quasitopological Yang-Mills solutions are accessible only for the positive values of $\Lambda$ and $\mu_{4}$. For space dimensions greater than eight, the solutions may have two black hole and cosmological horizons, if we consider $k=-1$ and $k=+1$ for respectively AdS and dS black holes. Finally depending on the values of the parameters $n$, $m$, $\Lambda$, $\mu_{4}$ and $e$, the solutions may lead to a black hole with two horizons, an extreme black hole or a naked singularity. \\
In the future works, we would like to search for the other quantities such as shadow, quasinormal modes, thermodynamic geometry and central charge of the obtained quasitopological Yang-Mills black holes.
 
\section{Appendix}
\subsection{Coefficients of the quartic quasitopological gravity} \label{app1}
The coefficients $b_{i}$'s and $c_{i}$'s for the cubic and quartic quasitopological terms in Eqs. \eqref{quasi3} and \eqref{quasi4} are respectively defined as
\begin{eqnarray}
&&b_{1}=3(3n-5)\nonumber\\
&&b_{2}=-24(n-1)\nonumber\\
&&b_{3}=24(n+1)\nonumber\\
&&b_{4}=48(n-1)\nonumber\\
&&b_{5}=-12(3n-1)\nonumber\\
&&b_{6}=3(n+1)\nonumber\\
\end{eqnarray}
and
\begin{eqnarray}
&&c_{1}=-(n-1)(n^7-3n^6-29n^5+170n^4-349n^3+348n^2-180n+36)\nonumber\\
&&c_{2}=-4(n-3)(2n^6-20n^5+65n^4-81n^3+13n^2+45n-18)\nonumber\\
&&c_{3}=-64(n-1)(3n^2-8n+3)(n^2-3n+3)\nonumber\\
&&c_{4}=-(n^8-6n^7+12n^6-22n^5+114n^4-345n^3+468n^2-270n+54)\nonumber\\
&&c_{5}=16(n-1)(10n^4-51n^3+93n^2-72n+18)\nonumber\\
&&c_{6}=-32(n-1)^2(n-3)^2(3n^2-8n+3)\nonumber\\
&&c_{7}=64(n-2)(n-1)^2(4n^3-18n^2+27n-9)\nonumber\\
&&c_{8}=-96(n-1)(n-2)(2n^4-7n^3+4n^2+6n-3)\nonumber\\
&&c_{9}=16(n-1)^3(2n^4-26n^3+93n^2-117n+36)\nonumber\\
&&c_{10}=n^5-31n^4+168n^3-360n^2+330n-90\nonumber\\
&&c_{11}=2(6n^6-67n^5+311n^4-742n^3+936n^2-576n+126)\nonumber\\
&&c_{12}=8(7n^5-47n^4+121n^3-141n^2+63n-9)\nonumber\\
&&c_{13}=16n(n-1)(n-2)(n-3)(3n^2-8n+3)\nonumber\\
&&c_{14}=8(n-1)(n^7-4n^6-15n^5+122n^4-287n^3+297n^2-126n+18).\nonumber\\
\end{eqnarray}
\subsection{Gauge potentials for some gauge groups}\label{app2}
The gauge potentials of the groups $SO(3)$, $SO(2,1)$, $SO(4)$ and $SO(3,1)$ are described as follows:\\
For $SO(3)$ guage group with $k=1$ and $n=4$,
\begin{eqnarray}
C^{1}_{23}&=& C^{2}_{31}=C^{3}_{12}=-1\,\,,\,\, \gamma_{ab}=\mathrm{diag}(1,1,1)\nonumber\\
A_{\mu}^{(1)}&=& e\,(-\mathrm{cos}\, \phi\, d\theta+\mathrm{sin}\, \theta \,\mathrm{cos}\,\theta \,\mathrm{sin}\,\phi \,d\phi),\nonumber\\
A_{\mu}^{(2)}&=& -e\,(\mathrm{sin}\, \phi\, d\theta+\mathrm{sin}\, \theta \,\mathrm{cos}\,\theta \,\mathrm{cos}\,\phi \,d\phi),\nonumber\\
A_{\mu}^{(3)}&=& e\,\mathrm{sin}^{2}\, \theta \,d\phi,
\end{eqnarray}
for $SO(2,1)$ gauge group with $k=-1$ and $n=4$,
\begin{eqnarray}
C^{1}_{23}&=& C^{2}_{31}=-C^{3}_{12}=1\,\,,\,\,\gamma_{ab}=\mathrm{diag}(-1,-1,1)\nonumber\\
A_{\mu}^{(1)}&=& e\,(-\mathrm{cos}\, \phi\, d\theta+\mathrm{sinh}\, \theta \,\mathrm{cosh}\,\theta \,\mathrm{sin}\,\phi \,d\phi),\nonumber\\
A_{\mu}^{(2)}&=& -e\,(\mathrm{sin}\, \phi\, d\theta+\mathrm{sinh}\, \theta \,\mathrm{cosh}\,\theta \,\mathrm{cos}\,\phi \,d\phi),\nonumber\\
A_{\mu}^{(3)}&=& e\,\mathrm{sinh}^{2}\, \theta \,d\phi,
\end{eqnarray}
for $SO(4)$ gauge group with $k=1$ and $n=5$
\begin{eqnarray}
C^{1}_{24}&=& C^{1}_{35}=C^{2}_{41}=C^{2}_{36}=C^{3}_{51}=C^{3}_{62}=1,\nonumber\\
C^{4}_{56}&=& -C^{4}_{21}=C^{5}_{64}=-C^{5}_{31}=C^{6}_{45}=-C^{6}_{32}=1,\nonumber\\
\gamma_{ab}&=&\mathrm{diag}(1,1,1,1,1,1),
\end{eqnarray}
\begin{eqnarray}
A_{\mu}^{(1)}&=& -e\,(\mathrm{sin}\, \phi\,\mathrm{cos}\, \psi\, d\theta+\mathrm{sin}\, \theta \,\mathrm{cos}\,\theta \,(\mathrm{cos}\,\phi\, \mathrm{cos}\,\psi \,d\phi-\mathrm{sin}\,\phi\, \mathrm{sin}\,\psi \,d\psi))\nonumber\\
A_{\mu}^{(2)}&=& -e\,(\mathrm{sin}\, \phi\,\mathrm{sin}\, \psi\, d\theta+\mathrm{sin}\, \theta \,\mathrm{cos}\,\theta \,(\mathrm{cos}\,\phi\, \mathrm{sin}\,\psi \,d\phi+\mathrm{sin}\,\phi\, \mathrm{cos}\,\psi \,d\psi))\nonumber\\
A_{\mu}^{(3)}&=& -e\,(\mathrm{cos}\, \phi\, d\theta-\mathrm{sin}\, \theta \,\mathrm{cos}\,\theta \,\mathrm{sin}\,\phi \,d\phi)\nonumber\\
A_{\mu}^{(4)}&=& -e\,\mathrm{sin}^{2}\, \theta \,\mathrm{sin}^{2}\, \phi \,d\psi\nonumber\\
A_{\mu}^{(5)}&=& e\,\mathrm{sin}^{2}\, \theta\,(\mathrm{cos}\, \psi\, d\phi-\mathrm{sin}\, \phi \,\mathrm{cos}\,\phi \,\mathrm{sin}\,\psi \,d\psi)\\
A_{\mu}^{(6)}&=& e\,\mathrm{sin}^{2}\, \theta\,(\mathrm{sin}\, \psi\, d\phi+\mathrm{sin}\, \phi \,\mathrm{cos}\,\phi \,\mathrm{cos}\,\psi \,d\psi)
\end{eqnarray}
and for for $SO(3,1)$ gauge group with $k=-1$ and $n=5$, we have
\begin{eqnarray}
C^{1}_{24}&=& C^{1}_{35}=C^{2}_{41}=C^{2}_{36}=C^{3}_{51}=C^{3}_{62}=1\nonumber\\
C^{4}_{56}&=& C^{4}_{21}=C^{5}_{64}=C^{5}_{31}=C^{6}_{45}=C^{6}_{32}=1\nonumber\\
\gamma_{ab}&=&\mathrm{diag}(-1,-1,-1,1,1,1),
\end{eqnarray}
\begin{eqnarray}
A_{\mu}^{(1)}&=& -e\,(\mathrm{sin}\, \phi\,\mathrm{cos}\, \psi\, d\theta+\mathrm{sinh}\, \theta \,\mathrm{cosh}\,\theta \,(\mathrm{cos}\,\phi\, \mathrm{cos}\,\psi \,d\phi-\mathrm{sin}\,\phi\, \mathrm{sin}\,\psi \,d\psi)),\nonumber\\
A_{\mu}^{(2)}&=& -e\,(\mathrm{sin}\, \phi\,\mathrm{sin}\, \psi\, d\theta+\mathrm{sinh}\, \theta \,\mathrm{cosh}\,\theta \,(\mathrm{cos}\,\phi\, \mathrm{sin}\,\psi \,d\phi+\mathrm{sin}\,\phi\, \mathrm{cos}\,\psi \,d\psi)),\nonumber\\
A_{\mu}^{(3)}&=& -e\,(\mathrm{cos}\, \phi\, d\theta-\mathrm{sinh}\, \theta \,\mathrm{cosh}\,\theta \,\mathrm{sin}\,\phi \,d\phi),\nonumber\\
A_{\mu}^{(4)}&=& e\,\mathrm{sinh}^{2}\, \theta \,\mathrm{sin}^{2}\, \phi \,d\psi,\nonumber\\
A_{\mu}^{(5)}&=& -e\,\mathrm{sinh}^{2}\, \theta\,(\mathrm{cos}\, \psi\, d\phi-\mathrm{sin}\, \phi \,\mathrm{cos}\,\phi \,\mathrm{sin}\,\psi \,d\psi),\\
A_{\mu}^{(6)}&=& -e\,\mathrm{sinh}^{2}\, \theta\,(\mathrm{sin}\, \psi\, d\phi+\mathrm{sin}\, \phi \,\mathrm{cos}\,\phi \,\mathrm{cos}\,\psi \,d\psi).
\end{eqnarray}

\acknowledgments{This work is supported by Irainian National Science Foundation (INSF). F.N would like to thank physics department of Isfahan University of Technology for warm hospitality.}
%%%%%%%%%%%%%%%%%%%%%%%%%%%%%%%%%%%%%%%%%%%%%%%%%%%%%%%%%%%%%%%%%%

\end{document}